\documentclass{aa}

\usepackage{graphicx}
\usepackage{txfonts}
\usepackage[colorlinks=true]{hyperref}
\usepackage{xcolor,colortbl}
\usepackage{float}
\usepackage{adjustbox}
\usepackage{pdflscape}
\usepackage{lscape}
\usepackage{comment}
\usepackage{changepage}
\usepackage{geometry}
\usepackage{threeparttable}
\hypersetup{colorlinks=true,
    linkcolor=blue,
    urlcolor=blue,
    citecolor=blue}

\usepackage{makecell}
\makeatletter
\renewcommand*\aa@pageof{, page \thepage{} of \pageref*{LastPage}}
\makeatother
\usepackage{diagbox}
\usepackage{multirow}
\usepackage{graphicx} %package to manage images
\graphicspath{ {./images/} }

\usepackage[rightcaption]{sidecap}

\usepackage{wrapfig}
\usepackage{tabularx}
\usepackage[caption=false]{subfig}
\begin{document} 

   \title{Mid-Infrared diagnostics for identifying main sequence galaxies in the local Universe}

%   \subtitle{I. Overviewing the $\kappa$-mechanism}

   \author{C. Daoutis
          \inst{1,2,3}
          \and
          A. Zezas \inst{1,2,3}
          \and
          M. L. N. Ashby \inst{3}
          }
   
   % \author{Autor et al.
   %        }

   \institute{ Physics Department, and Institute of Theoretical and Computational Physics, University of Crete, 71003 Heraklion, Greece \\
         \email{cdaoutis@physics.uoc.gr}
         \and
             Institute of Astrophysics, Foundation for Research and Technology-Hellas, 71110 Heraklion, Greece
        \and
            Center for Astrophysics | Harvard \& Smithsonian, 60 Garden St., Cambridge, MA 02138, USA
            }

   \date{Received 31 April 2025 / Accepted 17 September 2025}

% \abstract{}{}{}{}{} 
% 5 {} token are mandatory

  \abstract
   {A galaxy's mid-infrared
    spectrum captures a significant amount of information about its internal conditions such as the radiation field strength and its star formation activity. It is also intricately connected to dust characteristics and it contains spectral lines that serve as crucial indicators for galaxy activity diagnostics. Thus, characterizing galaxies' mid-infrared spectra is highly constraining of their nature.}
  % aims
   {This project describes a diagnostic tool for identifying main-sequence star-forming galaxies in the local Universe using infrared dust emission features that are characteristic galaxy activity.}
   {A physically motivated sample of mock galaxy spectra has been generated to simulate the infrared emission of star-forming galaxies in the local Universe. Using this sample, we developed a diagnostic tool for identifying main-sequence star-forming galaxies with machine learning methods. Custom photometric bands were defined to target key dust emission features, including polycyclic aromatic hydrocarbons (PAHs) and the dust continuum. Specifically, three bands were selected to capture PAH emission peaks at 6.2\,$\mu$m, 7.7 and 8.6\,$\mu$m, and 11.3\,$\mu$m, along with one band that estimated the strength of the radiation field that illuminated the dust. This diagnostic was subsequently applied to observed galaxies to evaluate its effectiveness in real-world applications.}
   {Our diagnostic achieves high performance scores, accurately identifying 90.9\% of main-sequence star-forming galaxies in a sample of observed galaxies. Additionally, it demonstrates low contamination, with only 16.2\% of AGN galaxies being misidentified as star-forming based on our test sample.}
  % conclusions heading (optional)
   {It is possible to combine observational studies and stellar population synthesis (SPS) frameworks to generate physically motivated simulated samples of star forming galaxies that exhibit similar spectral properties as their observed counterparts. By strategically positioning custom photometric bands on the mid-infrared spectrum to target specific dust emission features, our diagnostic can extract valuable information without the need to measure specific emission lines. Although PAHs are sensitive indicators of star formation and interstellar medium (ISM) radiation hardness, PAH emission alone is insufficient for identifying main-sequence star-forming galaxies. Finally, we have developed a physically-motivated spectral library of main sequence star-forming galaxies spanning from ultraviolet to far-infrared wavelengths.}


   \maketitle

%
%-------------------------------------------------------------------

\section{Introduction}

Galaxies' spectral energy distributions (SEDs) are important tools in astrophysical research. SEDs offer a panchromatic view of the physical processes that govern these cosmic entities. Analyzing SEDs helps us gain insights into the stellar populations, star formation histories, dust content, and conditions of the interstellar medium (ISM) within galaxies. Additionally, SEDs provide essential constraints for galaxy formation models and evolution, enhancing our understanding of the complex interplay between dust and radiation fields.

A galaxy's infrared (IR) spectrum contains a wealth of information such as various spectral features that are sensitive to the activity of the host galaxy, such as dust emission in the form of continuum emission, emission features like polycyclic aromatic hydrocarbons (PAHs) and forbidden emission lines. PAHs are molecules primarily composed of carbon and hydrogen atoms, typically arranged in carbon rings. These molecules absorb ultraviolet (UV) light and re-emit it in the form of IR light through vibrational relaxation \citep{2004ApJ...613..986P}. Their presence is evident in star-forming galaxies' IR spectra, which have prominent PAH features at 3.3, 6.2, 7.7, 8.6, 11.3, 12.7, and 17.1\,$\mu$m \citep{1984A&A...137L...5L,1985ApJ...290L..25A,2004ApJS..154....1W,2004ApJS..154..199S,2008ARA&A..46..289T}. These are generally attributed to a variety of emission mechanisms. In particular, the 3.3\,$\mu$m feature arises from C-H stretching\,$\mu$m, the 6.2\,$\mu$m feature from C-C stretching, the 7.7\,$\mu$m feature from coupled C-C and C–H bending, the 8.6\,$\mu$m feature from C-H in-plane bending, and the 10-15\,$\mu$m feature from C-H out-of-plane bending \citep{2001A&A...370.1030H,2008ARA&A..46..289T}.

Therefore, studies of galaxy's IR SED are extremely important for deducing key galaxy characteristics, such as stellar masses, star formation histories (SFHs), and active galactic nucleus (AGN) activity. For instance, the presence of a strong radiation source, like an AGN, can be indicated by high-ionization potential forbidden emission lines \citep{1992ApJ...399..504S}. One specific example is the [\ion{Ne}{V}] 14.3\,$\mu$m line, which has an excitation potential of 97.1 eV and cannot be generated solely by stellar processes. Additionally, there is increasing evidence that PAH emission features are often suppressed or entirely absent in low-metallicity environments \citep{2000NewAR..44..249M,2005A&A...434..867G,2006ApJ...639..157W,2006A&A...446..877M,2024ApJ...974...20W} or in regions with intense radiation fields \citep{2014MNRAS.443.2766A,2022MNRAS.509.4256G}.

The deployment of space-based observatories
in recent decades, such as the Infrared Space Observatory (ISO), the Spitzer Space Telescope \cite[SST;][]{2004ApJS..154....1W}, and James Webb space telescope \citep[JWST;][]{2006SSRv..123..485G} has made it possible to observe galaxies over a wider range of IR wavelengths, allowing for the development of detailed activity diagnostics. These diagnostics are typically based on integrated fluxes measured for bright, ubiquitous emission lines, or on the equivalent widths (EWs) of dust complexes. Various activity diagnostic methods have been developed over the years, including those by \cite{1992ApJ...399..504S}, \cite{2000A&A...359..887L}, \cite{2007ApJ...656..148A}, and \cite{2007ApJ...654L..49S}, among others. All such methods aim to identify the hardness of the radiation source, such as an AGN, responsible for heating the dust and producing the IR emission in galaxies. However, accurately isolating the fluxes or EWs of the features of interest requires making correct assumptions about the underlying starlight or dust continuum emission, which can be a challenging task.

A wealth of photometry from all-sky, broad-band galaxy surveys such as the Wide-field Infrared Survey Explorer \cite[WISE;][]{2010AJ....140.1868W} has led to a new generation of photometry-based activity diagnostics \citep{2011ApJ...735..112J,2012MNRAS.426.3271M,2012ApJ...753...30S,2014RMxAA..50..255C,2018ApJS..234...23A,2023A&A...679A..76D}, providing a complementary view of dust emission features in galaxies. However, extracting sub-feature variations of extended dust emission features, including PAH bands, from broadband WISE photometry presents a significant challenge due to continuum dilution.

Advances in computing power have opened new avenues for analyzing and manipulating large data sets. One of the most popular techniques for data analysis is the use of machine learning algorithms. Machine learning algorithms have a significant advantage: they can handle large amounts of data, identifying connections and structures that would be extremely difficult and time consuming to detect using traditional methods. Additionally, our improved understanding of stellar and ISM physics has led to the development of advanced codes capable of accurately simulating galaxy SEDs \citep{2019A&A...622A.103B,2021ApJS..254...22J}. This capability becomes crucial in scenarios where there is a lack or absence of multi-wavelength data, as it allows for the precise characterization of a galaxy population. 

In this work, we introduce a new diagnostic tool for identifying main sequence (MS) star-forming galaxies using machine learning methods. To develop this tool, we simulated physically motivated galaxy SEDs to produce a representative sample of local MS star-forming galaxies representing the properties of galaxy populations based on published samples. Additionally, we introduce new custom photometric bands in the mid-IR range (5-22\,$\mu$m) that encompass key features of star-forming galaxies, such as PAH bands and emission from stochastically heated dust, which serves as normalization and is a proxy for the hardness of the radiation field heating the dust. The bands are selected to be accessible by a wide range of IR observatories such as Spitzer and JWST. 

This paper is organized as follows. In Sect. \ref{sec2} the data samples are presented. In Sect. \ref{diagnostic} we describe the algorithm we used to develop our diagnostic tool. In Sect. \ref{secrs} we introduce the performance metrics used to evaluate our diagnostic and present the results from the evaluation of the performance. Finally, in Sect. \ref{discuss} we discuss the potential limitations, compare our diagnostic with current classification methods and examine the performance on borderline activity cases (mix of star formation and AGN) such as the class of composite galaxies.

\section{Galaxy samples} \label{sec2}

\subsection{Observed galaxy spectra} \label{obs_spec}

The Spitzer's Infrared Spectrograph \citep[IRS;][]{2004ApJS..154...18H} observed galaxies across a wavelength span of 5-38\,$\mu$m through four distinct modules. Among these modules, two operated at low resolution (R$\sim$60-130), with one operating within the short wavelength range (SL mode; 5.3–14\,$\mu$m) and the other covering long wavelengths (LL mode; 14–38\,$\mu$m). The remaining two modules operated at high resolution (R$\sim$600), designated for short (SH mode; 10–19.5\,$\mu$m) and long (LH mode; 19–37\,$\mu$m) wavelengths, respectively.

In this work, we use publicly available, reduced and calibrated galaxy spectra from the Spitzer Heritage Archive ({SHA}\footnote[1]{\url{https://irsa.ipac.caltech.edu/applications/Spitzer/SHA/}}). In particular, we select low-resolution galaxy spectra covering at least the 5-23\,$\mu$m wavelength range. In the SHA there are spectra from extragalactic legacy data sets that provide reduced spectra from several galaxy surveys. In particular, from SHA we were interested in surveys that have observed normal MS star-forming galaxies \cite[e.g., the Spitzer-SDSS-GALEX Spectroscopic Survey (SSGSS);][]{2011ApJ...741...79O}, extreme cases of galaxies that can harbor star formation, i.e., ultra-luminous infrared galaxies (ULIRGs) \cite[e.g., the Great Observatories All-sky LIRG Survey (GOALS);][]{2009PASP..121..559A}, and blue compact dwarfs (BCDs) which are low-mass and low-metallicity galaxies experiencing intense bursts of star-formation. These samples were chosen because combined they cover a wide range of galaxy conditions where star formation can exist, thus making them excellent for validating whether our diagnostic can be implemented reliably on actual observed samples of galaxies and assessing how well it can perform for a wide range of galaxy conditions.

One widely used sample of galaxies is the SIRTF Infrared Nearby Galaxies Survey \citep[SINGS;][]{2003PASP..115..928K}, which is a well-characterized set of 75 nearby galaxies. It was designed to provide a comprehensive view of the infrared properties of galaxies across a wide range of morphological types, luminosities, and star formation environments. However, SINGS primarily focuses on nuclear regions, and its extremely close proximity poses challenges in drawing general conclusions about galaxies as it suffers from strong aperture effects. Furthermore, SINGS has a significant amount of AGNs as well as IR luminous galaxies \citep{2011ApJ...741...79O}. Given the aforementioned concerns and the fact that our target population in this project comprises normal MS star-forming galaxies, we have excluded SINGS galaxies from our analysis.

The Spitzer SDSS Statistical Spectroscopic Survey \citep[S5;][]{2008sptz.prop50568S} and the SSGSS both contain optically selected star-forming galaxies that are representative of the normal galaxies present in the local Universe. The S5 is an extension of the SSGSS in a narrower redshift range ($0.05<z<0.1$) but including a larger number of galaxies. The sample covers galaxies with star formation rates of $-1.2 \leq$ log$_{10}($SFR) $\leq 1.6$, stellar masses of $9.1 \leq$ log$_{10}$($M_{*}) \leq 11.5$, and metallicities of $8.2 \leq Z \leq8.8$.
%(SSGSS:291)

The total number of galaxies in the S5 sample is 223. As the S5 sample was focused primarily on star-forming galaxies we find that it only contains 13 AGN and 22 composite galaxies, while the vast majority, 186, are star-forming galaxies. These activity classifications are based on a two-dimensional diagnostic plot using the emission-line ratios of [\ion{O}{III}]/H$\beta$ against [\ion{N}{II}]/H$\alpha$ \citep[BPT diagram;][]{1981PASP...93....5B,2001ApJ...556..121K,2003MNRAS.346.1055K,2007MNRAS.382.1415S}.

In order to make sure that our diagnostic performs well in identifying the vast majority of star-forming galaxies (i.e., high completeness) but also does not confuse other classes of galaxies as star-forming (i.e., contamination), we need to supplement our sample of galaxies with AGN as test cases. For this reason, we add 25 AGN galaxies from the sample of \cite{1997ApJS..112..315H}. In addition to them we find AGN galaxies included in the Infrared Database of Extragalactic Observables from Spitzer sample \citep[IDEOS;][]{2016MNRAS.455.1796H,2022ApJS..259...37S}. The IDEOS sample encompasses 3361 galaxies with redshifts up to 6.42. After selecting IRS spectra with S/N > 5, at the 14\,$\mu$m continuum and demanding that galaxies have reliable activity classifications derived from the {HECATE}\footnote[2]{\url{https://hecate.ia.forth.gr}} catalog \citep{2021MNRAS.506.1896K}, we identify 42 AGNs. In total, we have 186 star-forming galaxies and 80 AGN. All the above AGN and star-forming galaxies are located in the local Universe ($z<0.1$).

\subsection{Simulated galaxy spectra} \label{simgalspec}

The surveys presented in the previous section provide observed mid-IR spectra for a large number of galaxies over a broad range of physical parameters. However, the number of observed galaxies is not adequate for training a machine learning algorithm efficiently. To overcome this limitation we attempt to produce a suite of simulated galaxy spectra that have the same photometric properties (and subsequently similar physical properties) as their real counterparts by using the Flexible Stellar Population Synthesis \citep[FSPS;][]{2009ApJ...699..486C}. This allows us to produce galaxy spectra that could naturally exist and are very likely to be found in the local Universe but have not been observed due to limitations such as low survey coverage or poor spectra quality. This spectral library serves as the foundation for training machine learning based diagnostic tools.

Population synthesis codes are powerful in the respect that they fit the broad-band SEDs with several physical parameters characterizing the star-formation history, metallicity, and the ISM of a galaxy \citep[e.g.,][]{2016MNRAS.462.1415C,2018MNRAS.480.4379C,2019A&A...622A.103B,2021ApJS..254...22J}. However, this power can also be a weakness when population synthesis codes are used as simulators for broad-band spectra, as we do here. The large number of available parameters combined with the wide range of values they can take leads to almost infinite combinations with the many of them possibly being completely unrealistic. 

To define a spectral sample that is representative of typical star-formation dominated (i.e., MS and green valley) galaxies in the local Universe, we take two actions. First, from the numerous parameters that the FSPS offers we isolate the relevant parameters that are primarily responsible for the stellar populations and IR emission that are characteristic of star-forming galaxies. These are: the attenuation, gas-phase and stellar metallicity, fraction of dust mass in the form of PAHs, intensity of the radiation field of the interstellar medium, ionization parameter of the gas cloud, the dust fraction heated by a high-intensity radiation field, and the star formation history (SFH) of the galaxy. Second, in order to obtain realistic SEDs, the values of these parameters are based on studies of star-forming galaxies in the local Universe. In our analysis we adopt the works of \cite{2007ApJ...663..866D}, \cite{2020ApJ...889..150A}, \cite{2019ARA&A..57..511K}, and \cite{2022MNRAS.512.4136C} that provide an excellent census for distributions of the dust fraction exposure, fraction of dust mass in the form of PAHs and average radiation field strength, ionization parameter, and metallicity respectively. Table \ref{tab:physpar} lists each FSPS parameter with its corresponding value range and its probability distribution in more detail. Figure \ref{fig:param_dist} presents the distributions of the SPS parameters introduced in Table \ref{tab:physpar}.

\begin{figure}[h]
  \resizebox{\hsize}{!}{\includegraphics{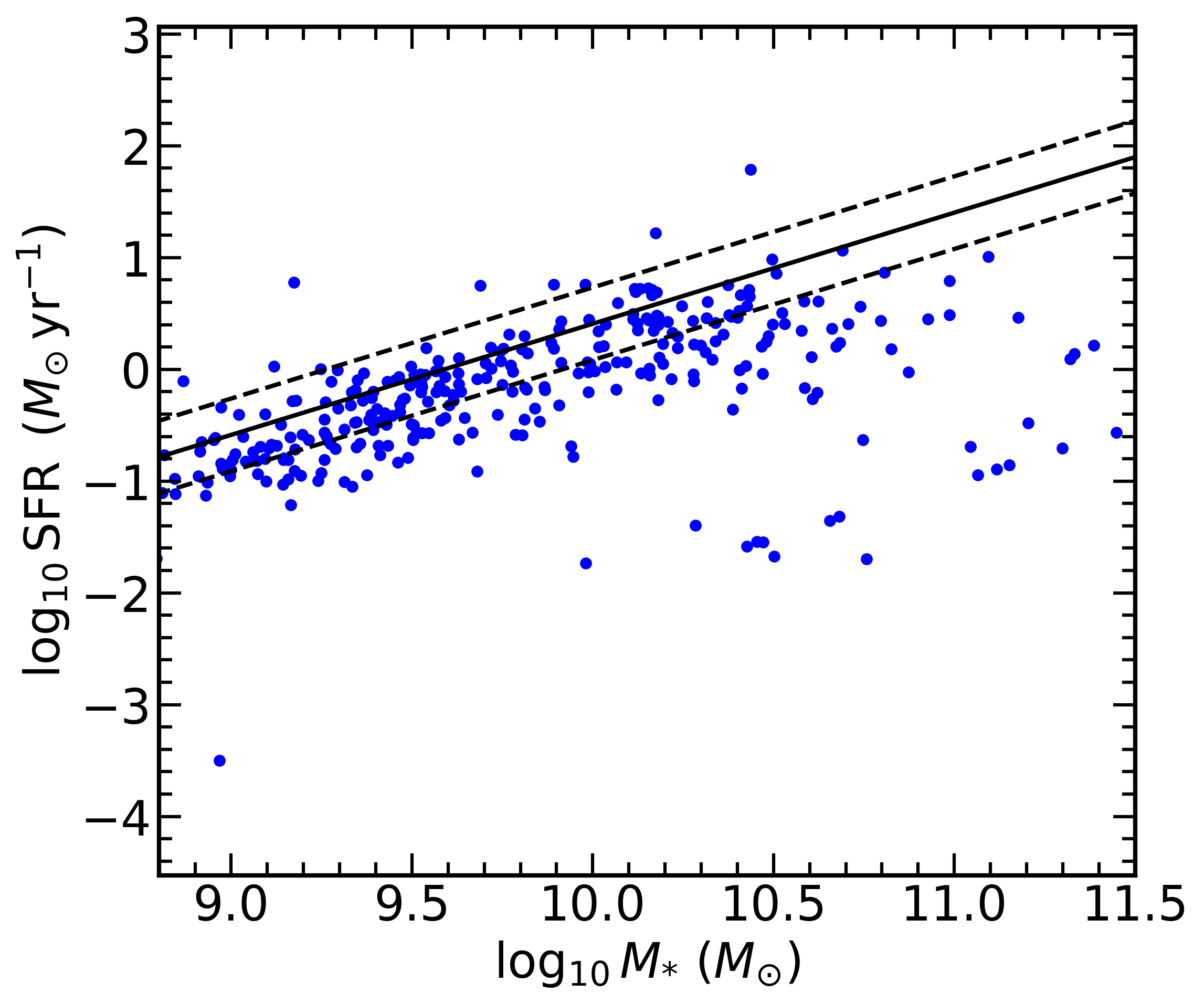}}
  \caption{Present-day SFR against stellar mass for the DustPedia galaxy sample. The black solid line represents the main sequence of galaxies as defined by \cite{2011A&A...533A.119E}, while the black dashed lines indicate the 1$\sigma$ deviations from this sequence. Most of the galaxies lie along the MS, with only a few falling below it.}
  \label{fig:SFR_M_star}
\end{figure}

All the parameter listed above are equally important for shaping the SED of a galaxy. However, SFH is arguably the most important, because it determines the spectral hardness
and the amount of radiation available to ionize the gas and heat the dust within a galaxy. For this reason, we selected SFHs from the {DustPedia}\footnote[3]{\url{http://dustpedia.astro.noa.gr}} sample \citep{2021A&A...649A..18G}. These SFHs were derived from SED fitting to a large number local Universe galaxies covering a wide range of SFH scenarios from starburst to passive galaxies. These galaxies also cover a wide range of stellar masses ($9<\text{log}_{10}M_{*}<11$) and SFRs ($-1.3<\text{log}_{10}\text{SFR}<0.5$). Thus, we consider that the wide variety of these SFHs adequately describes the typical star-forming galaxy population in the local Universe. Because we are interested in galaxies with infrared spectra dominated by stellar processes and in order to minimize biases in the SFH due to the AGN emission, we exclude from our sample galaxies that most likely host an AGN, based on mid-IR bands \citep{2019A&A...624A..80N}. Characterizing the activity of these galaxies with optical emission line diagnostics will lead to biases since their close distances make them prone to aperture effects. This results in a final sample of SFHs for 797 distinct galaxies (derived from real galaxies), encompassing a wide range of systems found in the star-forming MS. Figure \ref{fig:SFR_M_star} shows the relationship between the current SFR and stellar mass for the DustPedia sample, showcasing the SFR ranges that will serve as the foundation for our simulated sample. Most of the galaxies in that plot belong to the MS, with only a few objects below it. These galaxies are passively evolving but still have residual star formation.

\begin{table*}[ht]
\caption{Description, range, probability distribution function, and reference of each physical parameter we used to produce the simulated sample of galaxies.}
\centering
\begin{tabular}{l c c c }
\hline\hline
Description & Distribution & Distribution Variables \\
\hline
Metallicity (log$_{10} Z/Z_{\odot}$)\footnotemark[1] & Normal & $\mu=-0.1, \sigma=0.1$ \\
Ionization parameter (log$_{10}U$)\footnotemark[2] & Normal & $\mu=-3.0, \sigma=0.1$ \\
PAH-to-dust mass fraction ($q_{\mathrm{PAH}}$)\footnotemark[3] & Normal & $\mu= 3.35, \sigma= 1.2$ \\
Dust fraction exposure ($\gamma$)\footnotemark[4] & \makecell[l]{$f(x|k, \theta) = \frac{1}{\Gamma(k)\theta^k} x^{k-1} e^{-\frac{x}{\theta}}$} & $k = 0.97, \theta = 4.12$ \\
Interstellar radiation field strength ($U_{\mathrm{min}}$)\footnotemark[3] & \makecell[l]{$f(x|k, \theta) = \frac{1}{\Gamma(k)\theta^k} x^{k-1} e^{-\frac{x}{\theta}}$} & $k = 0.85, \theta = 0.039$ \\
\hline
\end{tabular}
\label{tab:physpar}
\tablefoot{Reference for each distribution:
\footnotemark[1]{\cite{2022MNRAS.512.4136C}},
\footnotemark[2]{\cite{2019ARA&A..57..511K}},
\footnotemark[3]{\cite{2020ApJ...889..150A}}, and
\footnotemark[4]{\cite{2007ApJ...663..866D}}.}
\end{table*}

To produce the sample of the simulated galaxy spectra we use the SFHs (derived from real galaxies used in the DestPedia project) as the backbone of this process. For each SFH we produce 1000 mock spectra by drawing each of the other five SPS parameters described earlier from the probability distribution functions described in Table \ref{tab:physpar} and presented in Fig. \ref{fig:param_dist}. This way each simulated spectrum is characterized by a SFH and a vector of SPS parameters drawn from observation-driven distributions. The SFH we adopted here comprises two distinct stellar components: a young stellar population and an old stellar population \citep{2016A&A...585A..43C,2017A&A...608A..41C}. These two populations are roughly indicative of a burst or quench of star formation, as well as a more passively evolving stellar component \citep{2019A&A...624A..80N}. For the attenuation law, we assume a starburst curve, as proposed by \cite{2000ApJ...533..682C} with a power-law modification introduced by \cite{2002ApJS..140..303L}. It is important to emphasize that each set of parameters are drawn randomly and independently (not from their joint five-dimensional space), since there are no systematic studies presenting the joint distributions of all these parameters this is the only available avenue, to avoid any biases.

{Our spectra are generated by assuming each time a SFH and varying five parameters, three regarding the dust (i.e., $\gamma$, $U_{\rm min}$, and $q_{\rm PAH}$) and two the stellar populations (metallicity and ionization parameter). To assess the variability within the mid-infrared spectrum that each dust parameter has, simulated spectra are generated, wherein each dust parameter is allowed to fluctuate within the 95\% of its specified probability function (Table \ref{tab:physpar}) while the other dust parameters are drawn from their probability distribution functions as presented in Table \ref{tab:physpar}. It is evident that the three parameters associated with dust models-$q_{\mathrm{PAH}}$, $\gamma$, and $U_{\mathrm{min}}$-have a more significant influence on the mid-infrared spectrum (Fig. \ref{fig:var_params_dust}) than the ones related to stellar (stellar metallicity) and gas (ionization parameter) (Fig. \ref{fig:var_params_gal}). For more details about the specifics of this exercise, please refer to Appendix \ref{Appnxb}.

\begin{figure*}[h]
\begin{center}
\includegraphics[scale=0.6]{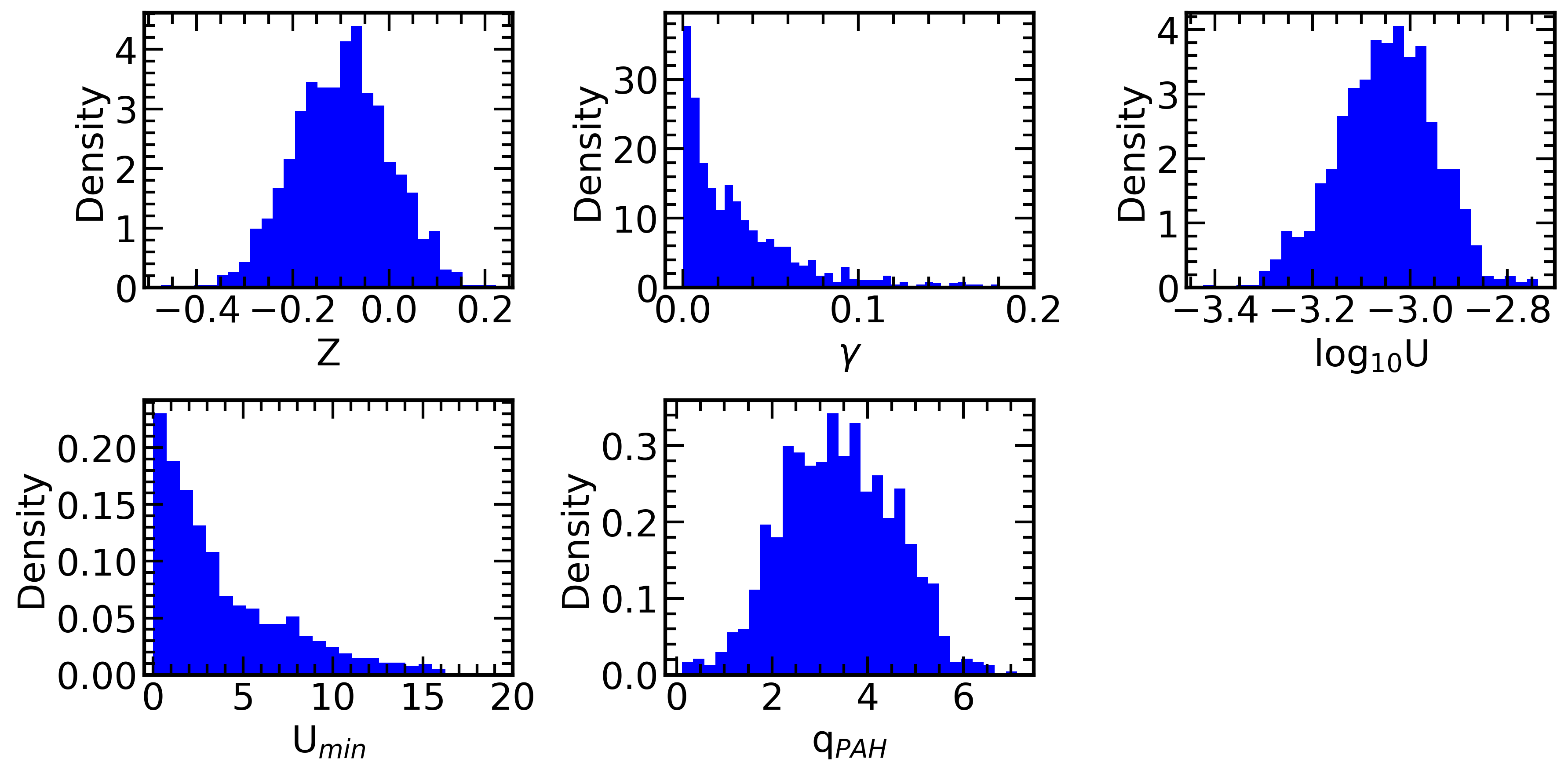}
\end{center}
    \caption{Histograms showing the shapes and the ranges of the distributions of the SPS parameters (Table \ref{tab:physpar}) used as an input to the FSPS to produce our mock sample of galaxies. In the top row, from left to right we show the distribution densities of metallicity (Z), the dust fraction exposed to a high ionization field ($\gamma$), and the ionization parameter of the gas (log$_{10}$U) respectively. On the bottom row on the left we show the minimum (average) ionization filed of the ISM (U$_{min}$) while on the left we see the fraction of dust in the from of PAH molecules (q$_{PAH}$). Each parameter set contains a set of these five parameters drawn randomly and independently from each other based on the probability distributions shown above.}
\label{fig:param_dist}
\end{figure*}

\subsection{Photometric properties of the simulated galaxy sample}

% In Section \ref{secftrsid}, we find that there is very good agreement between the distribution of the discrimination features of the simulated galaxies with the ones we obtained for an observed sample of star-forming galaxies. However, these four bands only cover a limited portion of the galaxy's SED and they do not provide information about how the simulated SEDs agree with the MS galaxy population in the rest of the photometry space. For this reason we use the color-color plots from ultraviolet (UV) to mid-IR.

To ensure that our simulated sample accurately reflects the photometric properties of an observed one, we construct and analyze color-color diagrams of $NUV-r$ against $g-r$, $g-r$ against $u-g$, and $W1-W2$ against $W2-W3$. By plotting these diagrams, we can visually assess the similarities and discrepancies between the simulated data and the actual observations. This process is crucial for validating the fidelity of our simulations in replicating the observed characteristics of galaxies.

We utilized the NUV photometric band from the Galaxy Evolution Explorer \citep[GALEX;][]{2005ApJ...619L...1M}, optical photometry in the u, g, and r bands from the Sloan Digital Sky Survey (SDSS), and the mid-IR W1, W2, and W3 bands from the WISE survey. In order to perform a more thorough comparison between our simulated SED sample and the overall galaxy population we use for the optical photometry the SDSS \citep[MPA-JHU;][]{2003MNRAS.346.1055K,2004MNRAS.351.1151B,2004ApJ...613..898T} catalog. We then cross-matched this catalog with the WISE All-sky catalog and the GALEX-SDSS-WISE Legacy Catalog \citep[GSWLC;][]{2016ApJS..227....2S} to acquire WISE and GALEX photometry respectively. In order to select star-forming galaxies from this galaxy sample, we used the diagnostic developed by \cite{2019MNRAS.485.1085S}. This diagnostic is based on the joint distribution of the four emission-line ratios: log$_{10}$([\ion{O}{III}]/H$\beta$), log$_{10}$([\ion{N}{II}]/H$\alpha$), log$_{10}$([\ion{S}{II}]/H$\alpha$), and log$_{10}$([\ion{O}{I}]/H$\alpha$). To ensure reliable classification with the \cite{2019MNRAS.485.1085S} diagnostic we imposed a strict selection criterion (S/N > 5 in all emission lines) even for weak emission lines like the [\ion{O}{I}] 6300\,$\AA$. This limits the selected population of the star-forming galaxies to be in the upper limit of the MS (i.e., intense star formation activity). In order to include in our comparison less actively star-forming galaxies that have weaker emission lines, or even galaxies without star-forming activity (passive galaxies), we perform photometric selection also based on the MPA-JHU sample and using the $NUV-r$ versus $M_{r}$ colour magnitude diagram \citep{2008MNRAS.385.1201H}. Subsequently, we calculate the $NUV-r$, $u-g$, $g-r$, $W1-W2$, and $W2-W3$ for our mock galaxy sample using the photometry derived from the simulated SEDs.

Figure \ref{fig:cc_NUV_r_g_r} presents the $NUV-r$ against $g-r$ color-color diagram for the simulated sample (black points) and the observed SDSS sample (red contours) of star-forming and passive galaxies (green contours). In that plot, our sample of simulated galaxies shows good overlap with the sample of the real galaxies in terms of color ranges and distribution, with only a slight deviation for bluer ($g-r$) colors and slightly redder ($NUV-r$) colors. Our simulated sample covers smoothly the transition from the bluer star-forming to the redder passive galaxies. However, we see the mock sample of galaxies does not cover the entire distribution of the observed passive galaxies. This is also expected behavior as at the redder end of the $NUV-r$ against $g-r$ color-color plot we have completely inactive galaxies with no star-formation processes, placing them outside the MS.

\begin{figure}[h]
\begin{center}
\includegraphics[scale=0.44]{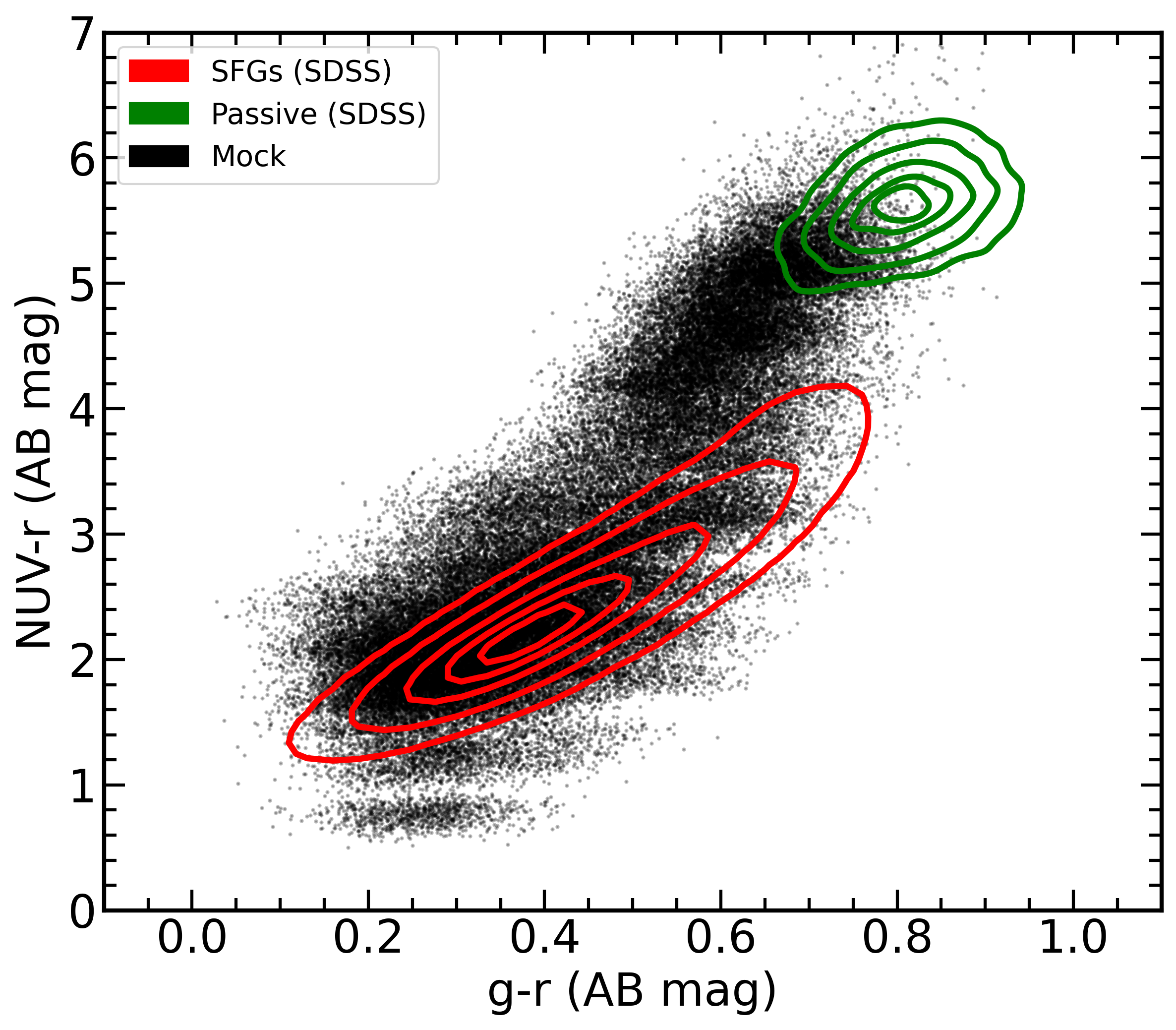}
\end{center}
\caption{Color-color diagram of $(NUV-r)$ against $(g-r)$ comparing the distributions of the mock (black dots) and a real SDSS (red contours) galaxy sample. We see a general agreement between the two galaxy samples for almost the entire range of these colors especially for $(g-r)$ $\sim$ 0.6. Photometric bands from GALEX (NUV) and SDSS survey (g and r). We also show a sample of photometrically selected passive galaxies (green contours) for comparison. The mock data were artificially augmented with noise to achieve S/N of 10 to match to the selection criteria for SDSS and GALEX bands. Contour levels denote the 10th, 25th, 50th, 75th, and 90th percentiles for each population.}
\label{fig:cc_NUV_r_g_r}
\end{figure}

Similar behavior we see in the color-color diagram of ($g-r$) against ($u-g$), which is shown in Fig. \ref{fig:cc_g_r_u_g}. The mock sample follows well the distribution of the SDSS galaxies with the only differences being the slightly redder ($u-g$) and bluer ($g-r$) colors, and a lack of the reddest and most passive galaxies. The difference in the bluest/UV colors could be the result of differences in the extinction law assumed in the simulations and the actual attenuation curve in real galaxies which can exhibit diversity from galaxy to galaxy \citep{2020ARA&A..58..529S}. Furthermore, this discrepancy may be attributed to the limitations of the emission line grids employed in SED frameworks, which typically only consider emission lines from very young stars ($\lesssim$ 10 Myrs). This restriction significantly impacts the H$\alpha$ emission and thus the r-band SDSS magnitudes. The colors we examined above are related to the stellar populations of a galaxy. To ensure that our mock sample of galaxies has dust emission properties similar to that of real galaxies, we also compare their infrared colors. In Fig. \ref{fig:cc_w1_w2_w3}, we plot the $W1-W2$, versus the $W2-W3$ colors calculated from the first three WISE bands. In the plot, we observe that the mock galaxies occupy the same range as the SDSS star-forming counterparts for both WISE colors. We also observe similar behavior for the passive galaxies. However, as in Fig. \ref{fig:cc_NUV_r_g_r} and Fig. \ref{fig:cc_g_r_u_g}, our mock galaxies do not cover the extreme end of the passive galaxies as expected.

\begin{figure}[h]
\begin{center}
\includegraphics[scale=0.45]{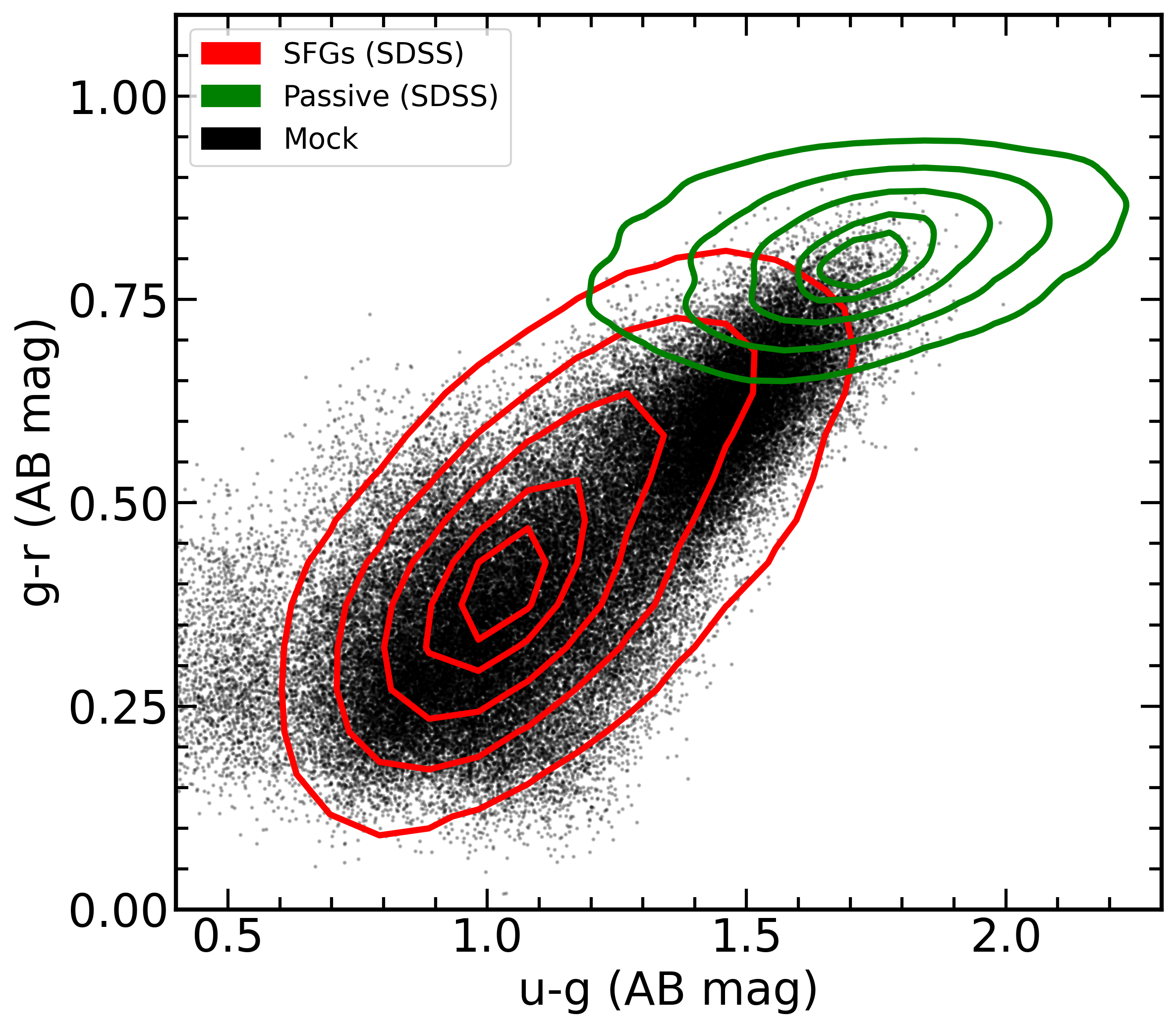}
\end{center}
\caption{Color-color diagram of $g-r$ against $u-g$ comparing the distributions of the mock (black dots) and a real SDSS (red contours) galaxy sample. We see a very good agreement between the two galaxy samples. Photometric bands from the SDSS survey. We also show a sample of photometrically selected passive galaxies (green contours) for comparison. The mock data were artificially augmented with noise to achieve S/N of 10 to match to the selection criteria for SDSS bands. Contour levels denote the 10th, 25th, 50th, 75th, and 90th percentiles for each population.}
\label{fig:cc_g_r_u_g}
\end{figure}

\begin{figure}[h]
\begin{center}
\includegraphics[scale=0.45]{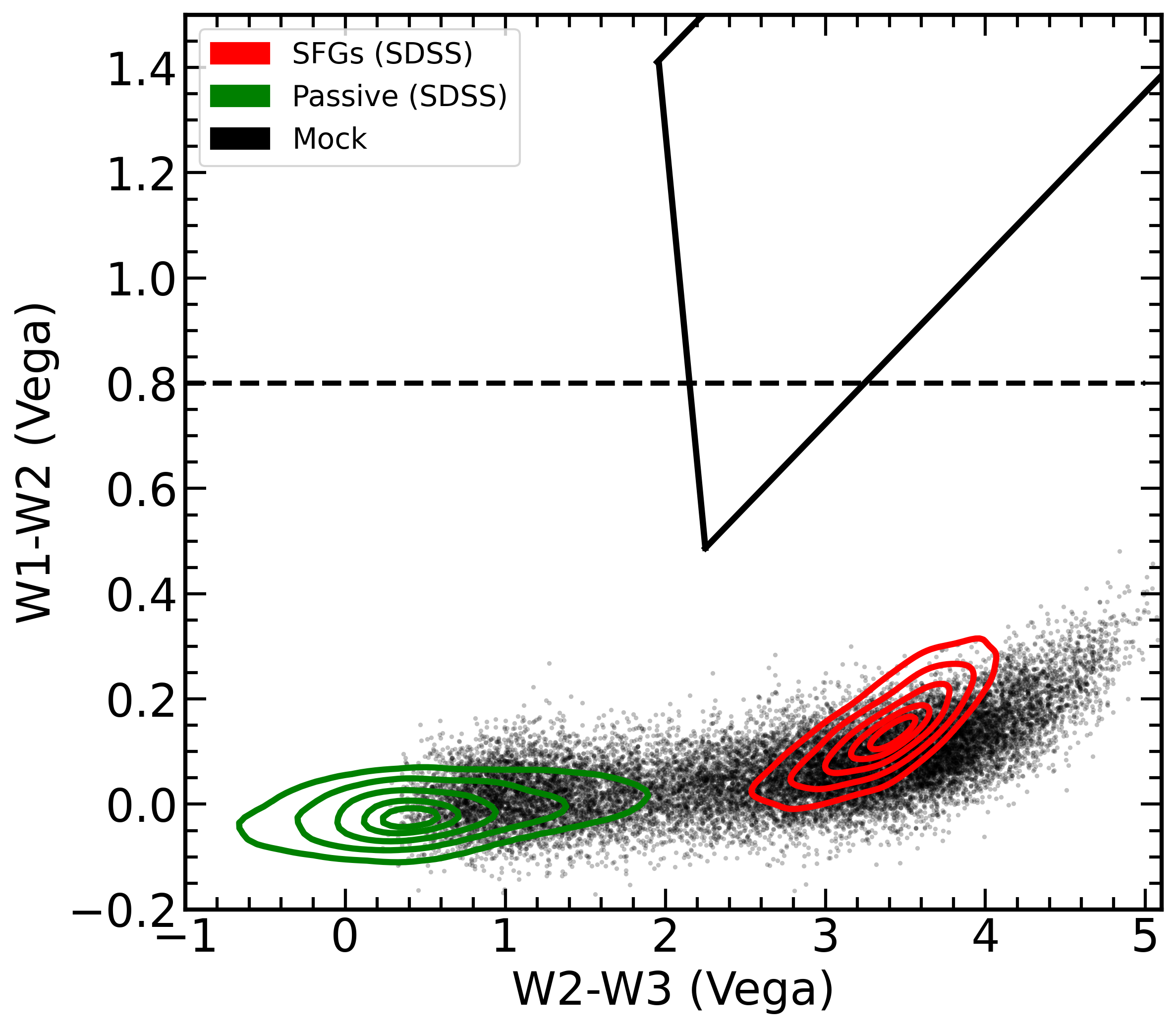}
\end{center}
\caption{Color-color diagram of $W1-W2$ against $W2-W3$ comparing the distributions of the mock (black dots) and a real star-forming galaxies from SDSS (red contours). We see a very good agreement between the two galaxy samples. Also both samples occupy the area well bellow the W1-W2 = 0.8 (black dashed line) the demarcation line defined by \cite{2012ApJ...753...30S} between AGN and non-AGN galaxies and the AGN selection wedge (solid black lines) defined by \cite{2012MNRAS.426.3271M}. We also show a sample of photometrically selected passive galaxies (green contours) for comparison. Photometric bands from the WISE survey. The mock data were artificially augmented with noise to achieve S/N of 10 to match to the selection criteria for WISE bands. Contour levels denote the 10th, 25th, 50th, 75th, and 90th percentiles for each population.}
\label{fig:cc_w1_w2_w3}
\end{figure}

All the results suggest that the mock SEDs have very good general agreement with the observed UV to mid-IR SEDs of galaxies. Additionally, the near and mid-IR emission in the mock sample of galaxies are in reasonable agreement with the simulations.

\section{The diagnostic tool} \label{diagnostic}

\subsection{Identifying the discriminating features} \label{secftrsid}

Galaxies' mid-IR spectra are rich with dust features and atomic lines that are directly linked to the underlying mechanisms heating the dust and gas. For example, the presence of high-ionization lines such as the [\ion{Ne}{V}] 14.3\,$\mu$m and [\ion{O}{IV}] 25.9\,$\mu$m (ionization potential of 97.1 eV and 54.9 eV respectively) is commonly regarded as a telltale sign of an AGN. However, these features require high-resolution spectra, which can be expensive to acquire, and sometimes also suffer from an additional complication in that their measurement involves subtracting the adjacent MIR continuum, especially in the presence of a bright AGN, or when the galaxy is unresolved. Another disadvantage is that to measure emission-line fluxes one must perform subtraction of the underlying continuum emission (i.e., stellar light and dust emission). This can often be a daunting task as the complex shape of the mid-IR part of the spectrum makes the estimation of the dust emission (i.e., PAHs) and absorption (e.g. silicates) features, very difficult. Even estimating the overall black-body continuum can be complicated by these broad band features if longer wavelength photometry is not available. 

% Diagnostic methods have been developed to exploit a wide spectral range, and by inference the ionization potentials of multiple lines, rather than a specific mission line, as a way to address the shortcomings associated with any single emission line. A galaxy is normally rich in spectral characteristics such as PAHs or parts of the spectrum indicative of the dust conditions. This can be achieved with photometric filters centered at specific areas of interest.

In order to facilitate the analysis of a large volume of data we focus of spectral bands encompassing diagnostically important features. We take care to avoid including multiple spectral features in each band in order to maximize their diagnostics power. 

We did not opt to use photometric filters from existing observatories such as Spitzer, WISE, or JWST because their broadband nature limits their ability to target specific PAH features. Although JWST includes three filters with central wavelengths and bandwidths comparable to our custom bands (i.e., F560W, F770W, and F1130W), its $\sim$21\,$\mu$m filter (F2100W) is too broad. It captures not only the underlying dust continuum but also blends PAH emission present in the 15–20\,$\mu$m, making it unsuitable for isolating the warm continuum emission.

Motivated by the previous works mentioned above, we began by incorporating features sensitive to the hardness of the radiation (i.e., [\ion{Ne}{V}] 14.3\,$\mu$m and [\ion{O}{IV}] 25.9\,$\mu$m) by defining bands centered on them. However, the underlying continuum proved to be strong enough to dilute any contribution from these emission lines, even in galaxies hosting AGN. Since our goal is to avoid modeling and removing the continuum, we focus on stronger spectral features that are associated with dust emission (i.e., PAHs). Specifically, we define three bands to include emission from PAHs targeting the peak wavelengths of 6.2, 7.7 and 8.6, and 11.3\,$\mu$m. 
The 12.7\,$\mu$m PAH feature was excluded from our selected feature scheme due to the substantial contribution from [\ion{Ne}{II}] 12.8\,$\mu$m, which significantly affects that spectral region. Additionally, we designate a band within a featureless (i.e., only continuum emission) part of the spectrum as a proxy for dust temperature and overall normalization. Centered at 21.5\,$\mu$m, this band serves as an indicator of the warm dust conditions which is stronger in AGN, and almost absent in passive galaxies. Table \ref{pah_bands} presents all the bands we defined, the wavelength coverage of each custom band, and the PAH features each one includes. We utilize three spectral bands (PAH1, PAH2, and PAH3) containing PAH features normalized by the continuum band (Cont215) as our discriminating features. After some experimentation, we chose to use the logarithm of these ratios to compress the data ranges and enhance the contribution of features with a small dynamic range.

\begin{table}[ht]
\caption{Wavelength range (rest frame) and important features that describe our custom fixed photometric bands.}
\centering
\begin{tabular}{l c c c c}
\hline\hline
Band name & Wavelength range ($\mu$m) &  PAH features ($\mu$m) & \\
\hline
PAH1 & 5.95-6.55 & 6.2\\
PAH2 & 7.20-8.90 & 7.7+8.6 \\
PAH3 & 10.90-11.81 & 11.3 \\
Cont215 & 21.00-22.00 & - \\
\hline
\end{tabular}
\label{pah_bands}
\end{table}

We chose the PAH features centered at 6.2, 7.7, 8.6, and 11.3\,$\mu$m as the basis of our feature scheme because there are no contamination from atomic spectral lines. Another advantage is their proximity to each other making them convenient to observe as they require only a short spectral range making this diagnostic widely applicable to a large number of galaxy samples from retired (e.g., Spitzer) to state-of-the-art (JWST) IR observing facilities. Figure \ref{fig:IRS_cutom_bands} shows an example spectrum of a star-forming galaxy (NGC 6120) observed with the IRS spectrograph. Figure \ref{fig:IRS_cutom_bands} demonstrates how the four fixed custom photometric bands we construct (Table \ref{pah_bands}) capture key spectral features characteristic of galaxy activity. In this plot, we see the location of the four custom photometric bands we chose to capture key spectral features that are characteristic of the activity of a galaxy. 

Having established the diagnostic bands and feature scheme, we proceed to measure the integrated fluxes within each such
band and divide the PAH by the continuum band. Fig. \ref{fig:dist_ftrs_real_mock} shows the distributions of PAH1, PAH2, and PAH3 normalized by the continuum at 21-22\,$\mu$m for the simulated spectral library. In order to verify that our simulated galaxy sample shares the same properties with the observed sample of real galaxies (Sect. \ref{obs_spec}), we compare these distributions between observed and mock galaxies. The distributions are very similar. To quantify this observation, we performed a Kolmogorov-Smirnov (K-S) test on the distribution of each discriminating feature and we found that the null hypothesis that the observed and mock distributions come from the same parent distribution cannot be rejected at a significance level higher than $\sim$15\% (p-value$\sim$0.85) for all three features.

\begin{figure}[h]
\begin{center}
\includegraphics[scale=0.4]{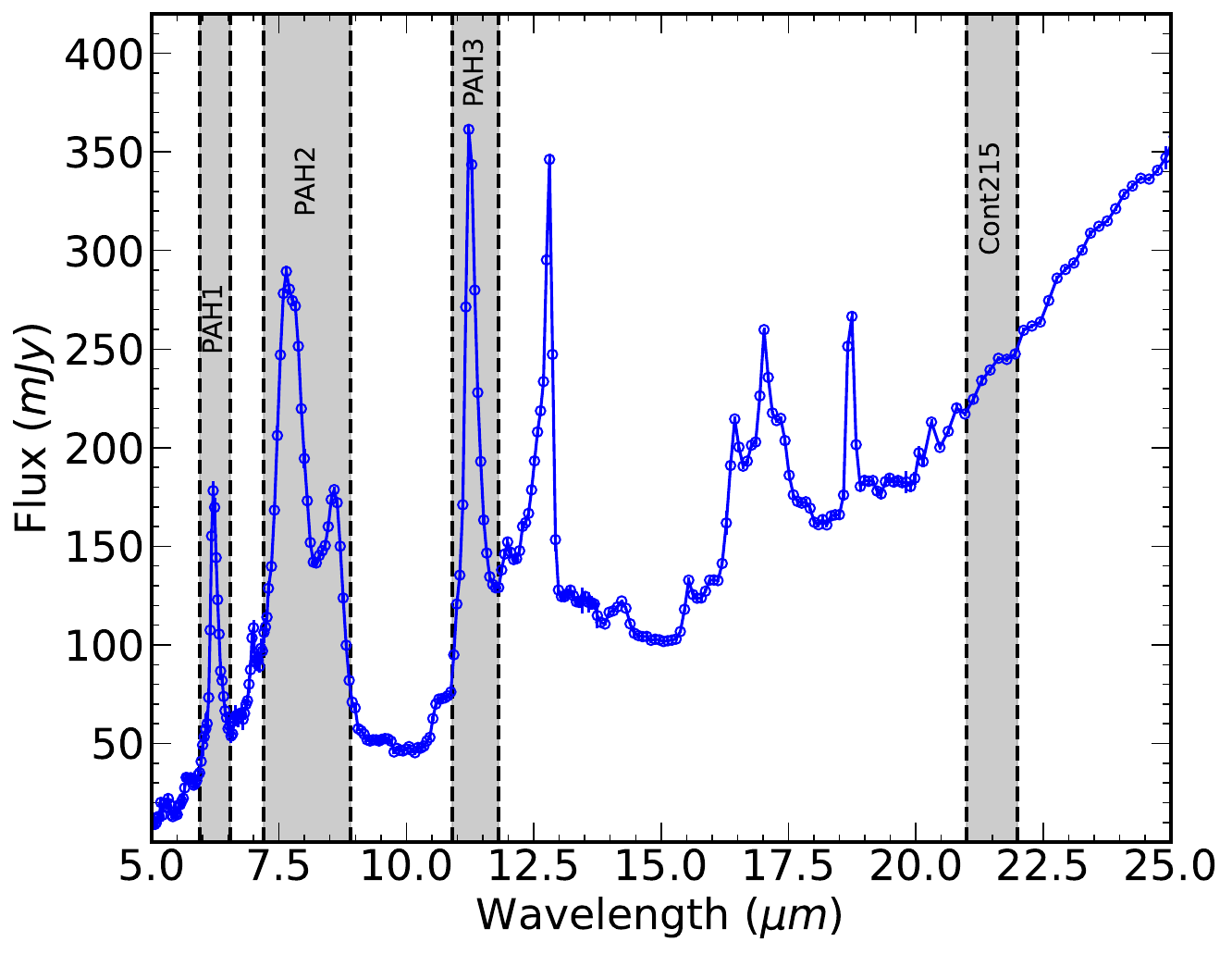}
\end{center}
\caption{The mid-infrared (5-25\,$\mu$m) spectrum of NGC 6120, observed with the IRS spectrograph onboard Spitzer, represents a typical star-forming galaxy (BPT class). In this spectrum, we have marked the wavelength coverage of our four custom photometric bands, labeled in Table \ref{pah_bands}. The first three photometric bands (PAH1, PAH2, and PAH3) cover the strongest PAH emission features, while the last band (Cont215) covers a continuum region of the spectrum.}
\label{fig:IRS_cutom_bands}
\end{figure}

\begin{figure}[h]
\begin{center}
\includegraphics[scale=0.5]{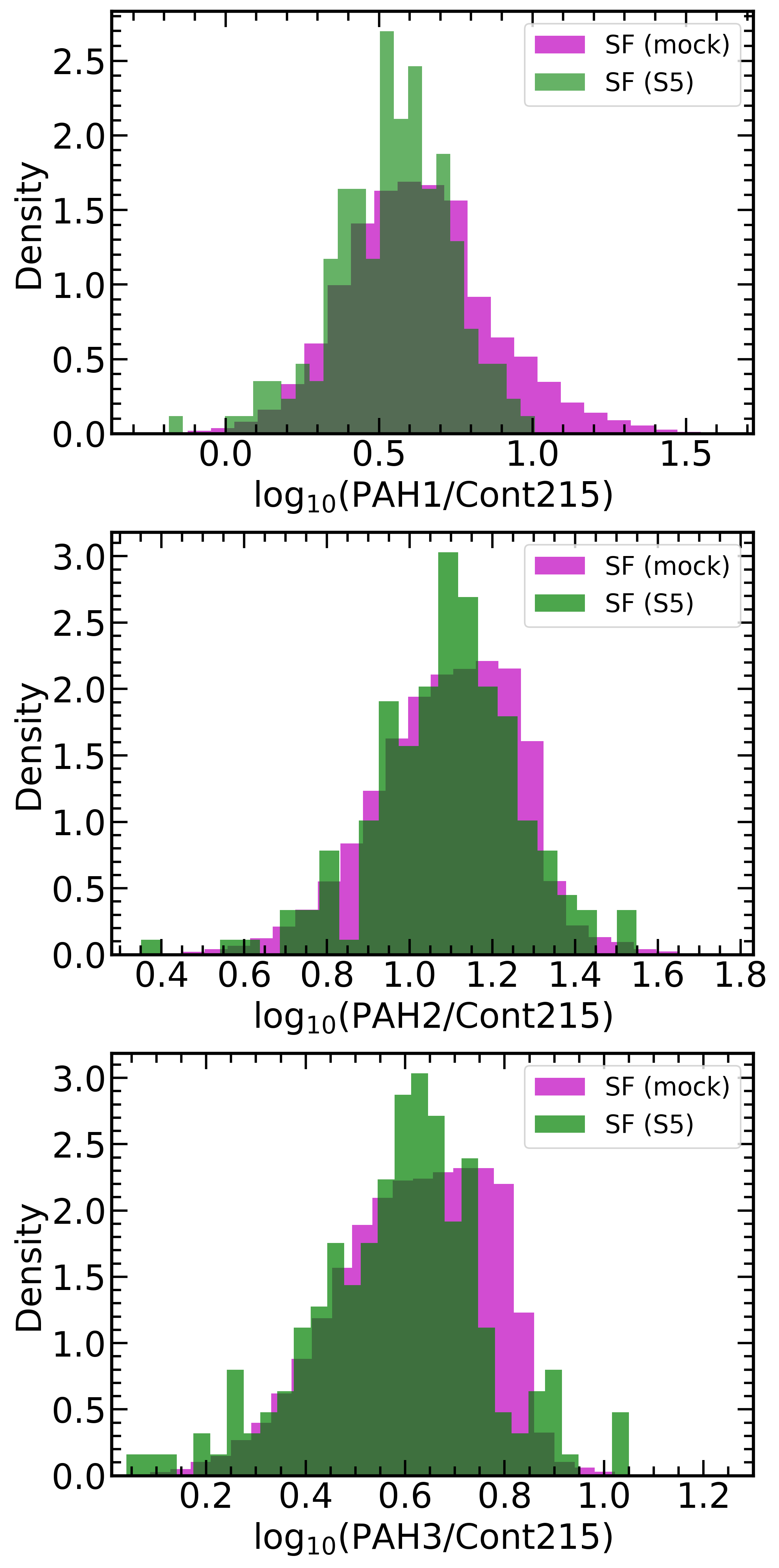}
\end{center}
\caption{Distributions of the three distinguishing features that were used to develop our diagnostic tool. From top to bottom, we see the log$_{10}$(PAH1/Cont215), log$_{10}$(PAH2/Cont215), and log$_{10}$(PAH3/Cont215), measured from both the S5 sample (green) of star-forming galaxies and our simulated sample of galaxies (purple). We found a very good agreement between the two samples for all the features.}
\label{fig:dist_ftrs_real_mock}
\end{figure}

\subsection{Choosing the algorithm}

There are several machine learning algorithms that can be applied in classification problems. Since our study is focused on the identification of a single type (class) of objects, that of population of star-forming galaxies, it is essential to utilize an algorithm capable of handling single-population data. In general, this kind of problem is handled by training an algorithm to learn the characteristics of a population in the selected feature space. After the training, during the classification process, any object that does not belong to the target population is flagged as an outlier. There is a wide variety of such algorithms ranging from more sophisticated (e.g., artificial neural networks) to simpler ones \citep[e.g., isolation forest;][]{4781136}. After testing various such algorithms, we found that the best performance is achieved by using a variant of the Support Vector Machine \citep[SVM;][]{cortes1995support}, specifically the one-class SVM (\texttt{OneclassSVM}) which is provided through the \texttt{scikit-learn} library based on the Python programming language.

\subsection{Training of the algorithm} \label{trlg}

During training, an SVM aims to find an optimal hyperplane that effectively separates the data points belonging to different classes in a high-dimensional feature space by adjusting the parameters of the hyperplane, guided by the labeled training data (i.e., supervised learning). In our case, the data describe only one galaxy population (star-forming galaxies). The \texttt{OneClassSVM} is an unsupervised algorithm that constructs an envelope around the data based on their distributions in the multidimensional feature space.

As mentioned in Sect. \ref{simgalspec} the spectra available for training our diagnostic are scarce. For this reason, the training of our diagnostic was conducted exclusively on simulated data, utilizing the library of simulated SEDs introduced in Sect. \ref{simgalspec}. This approach enables us to employ a large, representative sample of star-forming galaxies in the local Universe overcoming the limitations imposed by the limiting number of observed galaxies. We divided the simulated galaxy sample (Sect. \ref{simgalspec}) into two subsets: 70\% for model training and 30\% for performance testing.

Following the training phase, we evaluated the diagnostic's performance on both the simulated test set (which was not involved in the training) and a real galaxy sample (Sect. \ref{obs_spec}). This dual evaluation serves two purposes. First, it allows us to verify that performance for the training and test sets are consistent; discrepancies here could indicate overfitting or underfitting. Second, we assess the diagnostic’s effectiveness on real (observed) spectra to confirm its applicability and robustness in realistic conditions, ensuring its reliability on observed galaxy data.

\section{Results} \label{secrs}

\subsection{Performance metrics}

To assess the efficacy of our diagnostic, we commence by defining the classification objective: distinguishing between MS star-forming and other types of galaxies (including AGN, passive galaxies, or even extreme star-forming galaxies like ULIRGs or metal poor starbursts). We employed widely adopted metrics from supervised learning to quantify performance. Accuracy denotes the proportion of correctly classified galaxies out of the total sample, providing a comprehensive measure of the model’s effectiveness. A high accuracy indicates the correctness of the majority of classifications. Recall, conversely, quantifies the completeness of the identified class, in our context, the fraction of true star-forming galaxies that the model successfully identifies as such. A high recall score signifies the model’s ability to recover the entire target star-forming galaxy population, even if some misclassifications occur. Furthermore, we quantify the contamination from other classes, by recording the number of galaxies that were predicted to be MS star-forming galaxies by our diagnostic but were originally classified (ground truth) to belong to other activity classes (e.g., AGN and composite). Low contamination indicates a highly effective diagnostic tool.

\subsection{Evaluation of performance} \label{sec_eval_prf} 

After training and optimizing our diagnostic (see Appendix \ref{apndx1} for details), we evaluated its performance in two cases. First, we assessed its performance on the test set (see Sect. \ref{trlg}) and found it achieves 90\% recall, similar to its training set score, ensuring good generalization to similar data sets. When a diagnostic is designed to identify a single class, this evaluation primarily ensures good generalization to other datasets, preventing overfitting. However, it cannot provide information about misclassified objects that do not belong to the targeted class (i.e., contamination, see Appendix \ref{apndx1}). For the second test, we assess its performance on an observed sample of star-forming galaxies (Sect. \ref{obs_spec}). This test is essential to validate its relevance for real-world galaxy samples. Achieving a high recall score would not only demonstrate its effectiveness in identifying MS star-forming galaxies but also affirm its reliability for actual observations.

We selected two spectral samples of observed galaxies in order to test the performance of our diagnostic as a function of activity type (see Sect. \ref{secftrsid}). The measurements of the diagnostic features on the observed spectra were performed using the same photometric bands presented in Sect \ref{pah_bands}. The activity classification (i.e., the ground truth) of both samples is based on optical spectroscopy (BPT). The first contains MS star-forming galaxies in the local Universe. The application of our diagnostic on that sample should indicate how well it can identify the population of galaxies we are targeting. The other sample contains galaxies that host an active nucleus. AGN galaxies have generally weaker PAH and stronger mid-IR continuum than star-forming galaxies. Thus, our diagnostic should identify any galaxy that is not related with star formation processes as an abnormal instance and not as normal star-forming galaxy.

After applying our diagnostic to both samples of observed galaxies (star-forming and AGN) we achieve an overall accuracy of 88.7\%. Furthermore, we find that the recall score on the MS star-forming galaxies is 90.9\%. In addition, by applying our diagnostic on the sample of spectroscopically identified AGN galaxies we find that 16.2\% of AGNs have been classified as normal star-forming galaxies. Overall, the performance of our diagnostic is very high, and similar to the recall we achieved when we evaluated our diagnostic on the mock sample of galaxies. This indicates that our diagnostic can be reliably applied to real galaxies without loss in performance. In addition, the low contamination from AGN means that the boundaries that contain the star-forming galaxies in the selected feature space are well-defined, offering high recall and low contamination. In Sect. \ref{lmts} we make a more thorough analysis of the AGN population that contaminates our diagnostic results.

\subsection{Stability of the performance}

% As discussed in Sect. \ref{trlg} our diagnostic was trained on photometric data from simulated SED derived from observation-driven distributions of galaxy physical parameters. However, these samples do not include information about measurement uncertainties, which is an integral part of observational data. Therefore, it is important to ensure that our diagnostic performs well with noise. To achieve this, we conducted Monte Carlo simulations based on the signal-to-noise ratio of the measured discriminating features of the mock sample. These simulations helped us determine whether our scores are robust enough to withstand small perturbations caused by the quality of measurements.

In order to simulate the effect of different spectral quality, we measure the flux in each of the four custom photometric bands for the star-forming and AGN galaxies in the observed sample. Then, we calculate the values of the discriminating features. For each feature, we draw 1000 samples from a normal distribution. The mean of the distribution is the feature value, and the standard deviation is determined by the target S/N value. We consider data with S/N of 3, 5, 10, and 50. We apply our diagnostic on the sample of real AGN and star-forming galaxies individually and we record the recall of each activity class. for S/N values of 3, 5, 10, and 50. Our results demonstrate that the diagnostic performance is unaffected for both AGN and star-forming galaxies as the recall and the accuracy remained almost unchanged.

These results demonstrate the effectiveness and the robustness of our diagnostic tool in accurately identifying star-forming galaxies even when the measurements are of low quality. In particular, our tool is able to accurately identify MS star-forming galaxies and discriminate them from AGN with high recall and low contamination.

\section{Discussion} \label{discuss}

In this work, we produced a library of simulated SEDs for MS star-forming galaxies in the local Universe. The sample is representative in terms of covering a wide range of ISM physical parameters (e.g., the dust fraction exposed to
a high ionization field, the ionization parameter, the average ionization filed of the ISM, and the fraction of dust in the from of PAH molecules) and stellar populations (SFH and metallicity). Based on this library, we defined a diagnostic tool in order to discern star-forming galaxies from the rest of the galaxy activity types. In the following sections, we discuss into its behavior and its potential limitations.

\subsection{PAH emission features as galaxy activity probes} \label{s51}

In Sect. \ref{secrs}, we found that our diagnostic achieved excellent performance, even though it was provided with only three discriminating features. This follows from the well known sensitivity of PAHs of the interstellar radiation field and the PAH abundance as a result of star formation \citep{2007ApJ...656..770S}.

It has been observed that AGN-dominated galaxies typically exhibit suppressed PAH emission features, resulting in lower PAH EWs \citep{2007ApJ...656..148A} compared to star-forming galaxies. This suppression can be attributed to either the destruction of PAH molecules, which are sensitive to radiation strength \citep{1991MNRAS.248..606R,1992MNRAS.258..841V,2004A&A...414..123S,2022MNRAS.509.4256G} or the dilution of their emission in the strong AGN continuum \citep{2014MNRAS.443.2766A,2014MNRAS.445.1130R,2015MNRAS.449.1309G}.

However, even in galaxies with extreme conditions (i.e., starbursts or AGNs) where the PAH emission is normally expected to be severally depressed, PAH emission can still be detected. In other words, if we only rely on the presence or the absolute luminosity of PAH features we may misidentify the activity of these galaxies as being MS star-forming galaxies. To overcome this limitation, we incorporated an additional band (Cont215, see Table \ref{pah_bands}) as a normalization factor to enhance the discriminatory power of our diagnostic. This band is sensitive to the hardness of the radiation field that illuminates the dust. As the radiation field becomes harder, the slope at 21-22 $\mu$m increases due to the stochastically heated dust or obscuration affecting the continuum emission. Consequently, as the radiation strength of the field increases, we observe a decrease in PAH emission while simultaneously observing an increase in continuum emission at 21-22 $\mu$m. This way we can exclude galaxies that may have ongoing star formation under abnormal conditions (i.e., ULIRGs or low-luminosity AGN) reducing the contamination of non-MS star-forming galaxies in our diagnostic.

\subsection{Application on a sample of composite galaxies} \label{comp_av_spec}

In some cases, the activity of a galaxy can not be uniquely attributed only to star-formation processes or an active nucleus. In other words, a galaxy's SED may contain contributions by star-formation processes and the presence of an active nucleus. One notable example is the composite galaxies. Naturally, we expect these galaxies to be the transition between pure star-forming galaxies and AGNs. Thus, we wish to test our diagnostic on a sample of composite galaxies in order to assess its performance on galaxies with AGN and star-forming contributions. To do this we identify objects from the SSGSS/S5 survey that have been classified as composites based on their BPT classifications. Then we apply our diagnostic on these samples of galaxies to classify them. We find that 50\% are classified as star-forming while the other 50\% as not star-forming galaxies.

\begin{figure}[h]
\begin{center}
\includegraphics[scale=0.39]{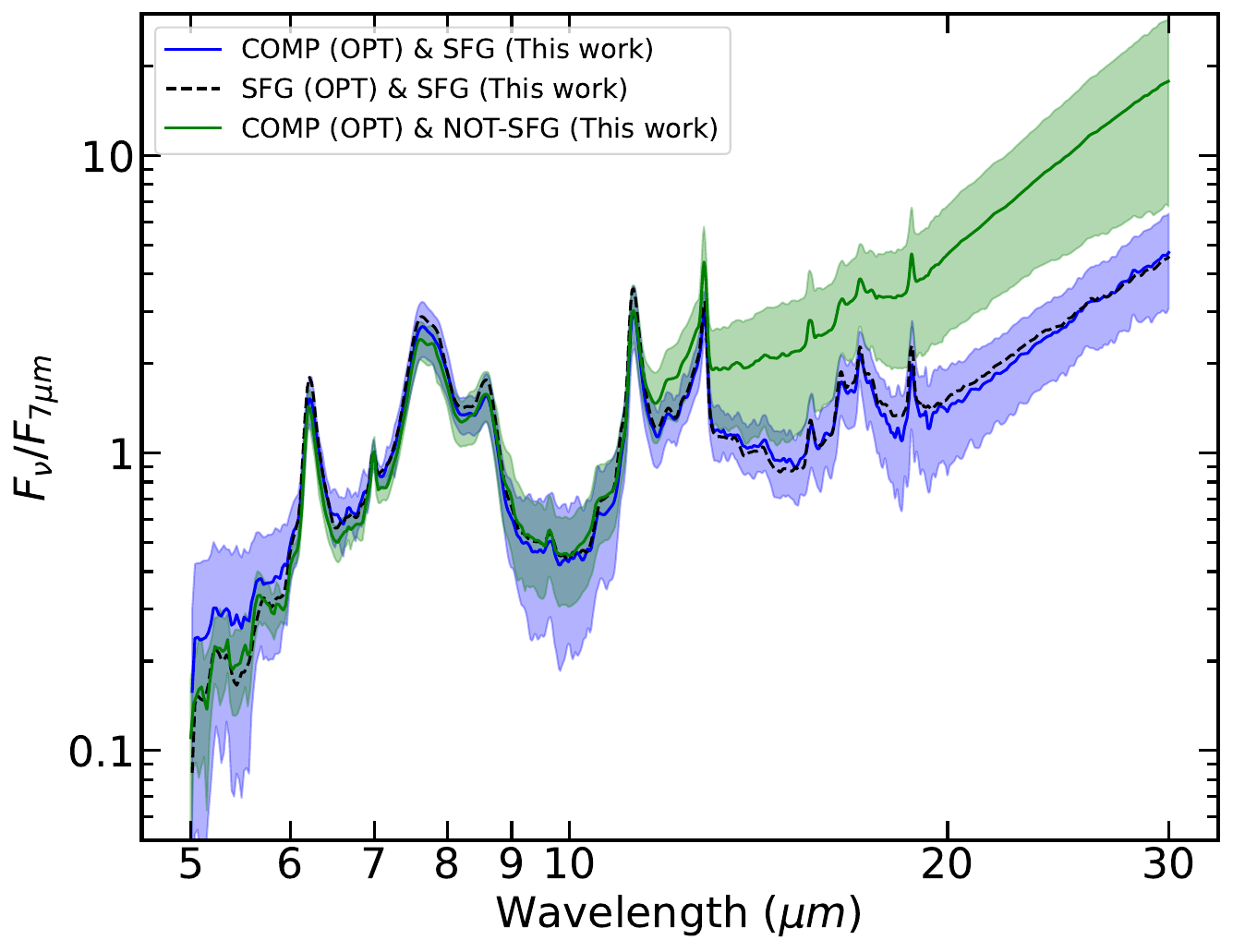}
\end{center}
\caption{Average mid-IR spectra of composite galaxies (based on the BPT optical diagnostics) classified as star-forming (SFG) and non-star-forming (NOT-SFG) based on our diagnostic. The composite galaxies predicted as star-forming galaxies (blue solid line) have similar average spectra with the typical MS star-forming galaxies (black dashed line). The composite galaxies that have been predicted as not star-forming (blue solid line) have steeper spectra above 12\,$\mu$m. The shaded area in both average spectra corresponds to the standard deviation of 1$\sigma$.}
\label{fig:mn_SED_comp}
\end{figure}

Afterward, in order to understand the driver of these results we calculate the average mid-IR spectrum of the galaxies that are classified as star-forming and non-star-forming based on our diagnostic. The average mid-IR spectra are calculated by stacking the individual (rest-frame) mid-IR spectrum for each class after normalizing them  at 7\,$\mu$m. Figure \ref{fig:mn_SED_comp} shows the average mid-IR spectrum of composite galaxies (BPT classification) that have been predicted as being star-forming and as not star-forming (our diagnostic). It is clear that in the spectral range of 5-13\,$\mu$m the PAH features are prominent in both subpopulations making them almost indistinguishable. On the other hand the average mid-IR spectrum of the composite galaxies that have been predicted as star-forming have significantly lower fluxes in the spectral range above 13\,$\mu$m. This result indicates that the composite galaxies that have been classified as not resembling a normal star-forming galaxy have dust stronger hot dust emission.

Furthermore, after placing these optically classified composite galaxies on the BPT diagram of [\ion{O}{III}]/H$\beta$ versus [\ion{N}{II}]/H$\alpha$, we have not observed any trend for the composite galaxies that our diagnostic identified as star-forming and the ones that were identified as non-star-forming galaxies. All of these observations can be attributed to the fact that dust may hinder our ability to identify an obscured AGN in the optical due to extinction, as suggested by \citep{2009MNRAS.398.1165G,2018ARA&A..56..625H}.

\subsection{Physical constraints} \label{lmts}

In Sect. \ref{sec_eval_prf} we tested the performance of our diagnostic on a sample of MS star-forming galaxies. However, star formation is also present in galaxies that are away from the main sequence. For example, ULIRGS have total IR luminosities exceeding 10$^{11} L_{\odot}$ and are often associated with ongoing galaxy mergers or interactions which can trigger intense star-formation (starburst galaxies) or supply material to the nuclear supermassive black hole of the merging galaxies leading to the formation of an AGN. 

To assess the performance of our diagnostic in such extreme galaxies, we applied it to galaxies from the GOALS sample. GOALS \citep{2009PASP..121..559A} includes 189 nearby galaxies with redshifts ranging from 0.003 to 0.09. Only  three galaxies were classified as main-sequence star-forming galaxies by our diagnostic, consistent with our expectations since almost none of them belong to the MS.

The other extreme end in galaxy populations that can harbor star formation are the BCDs. These extremely small-sized galaxies have stellar masses of $10^{7}-10^{9}M_{\odot}$ and typically have high specific star-formation rates (sSFR). To assess the behavior of our diagnostic on that extreme type of galaxies we used the sample of BCDs of \cite{2019ApJ...884..136X}. We find that none of the BCDs was classified as MS star-forming galaxies according to our diagnostic. This is driven by two effects. The first is that BCDs have suppressed PAH emission \citep{2004ApJ...613..986P,2005ApJ...628L..29E,2006A&A...446..877M,2006ApJ...639..157W,2016ApJ...818...60S} in their IR spectra. This could be a result of the low metallicities of BCDs leading to low dust content \citep{2007ApJ...663..866D}. The second is that the intense star formation in combination with their low metallicities create a strong UV radiation field which in combination with the low optical depth of the dust results to stochastic heating of the small dust grains contributing to the mid-IR part of the spectrum \citep{2003MNRAS.343..839T}.

In Sect. \ref{sec_eval_prf} while testing the performance of our diagnostic we found that 16.2\% of the AGN sample (based on optical spectral classifications) were classified as star-forming based on our diagnostic. To analyze this further we calculate the average spectra of the AGN galaxies (true class) that are classified as star-forming and non star-forming based on our diagnostic. As in the case of the composite galaxies the spectra are normalized at 7\,$\mu$m.

\begin{figure}[h]
\begin{center}
\includegraphics[scale=0.39]{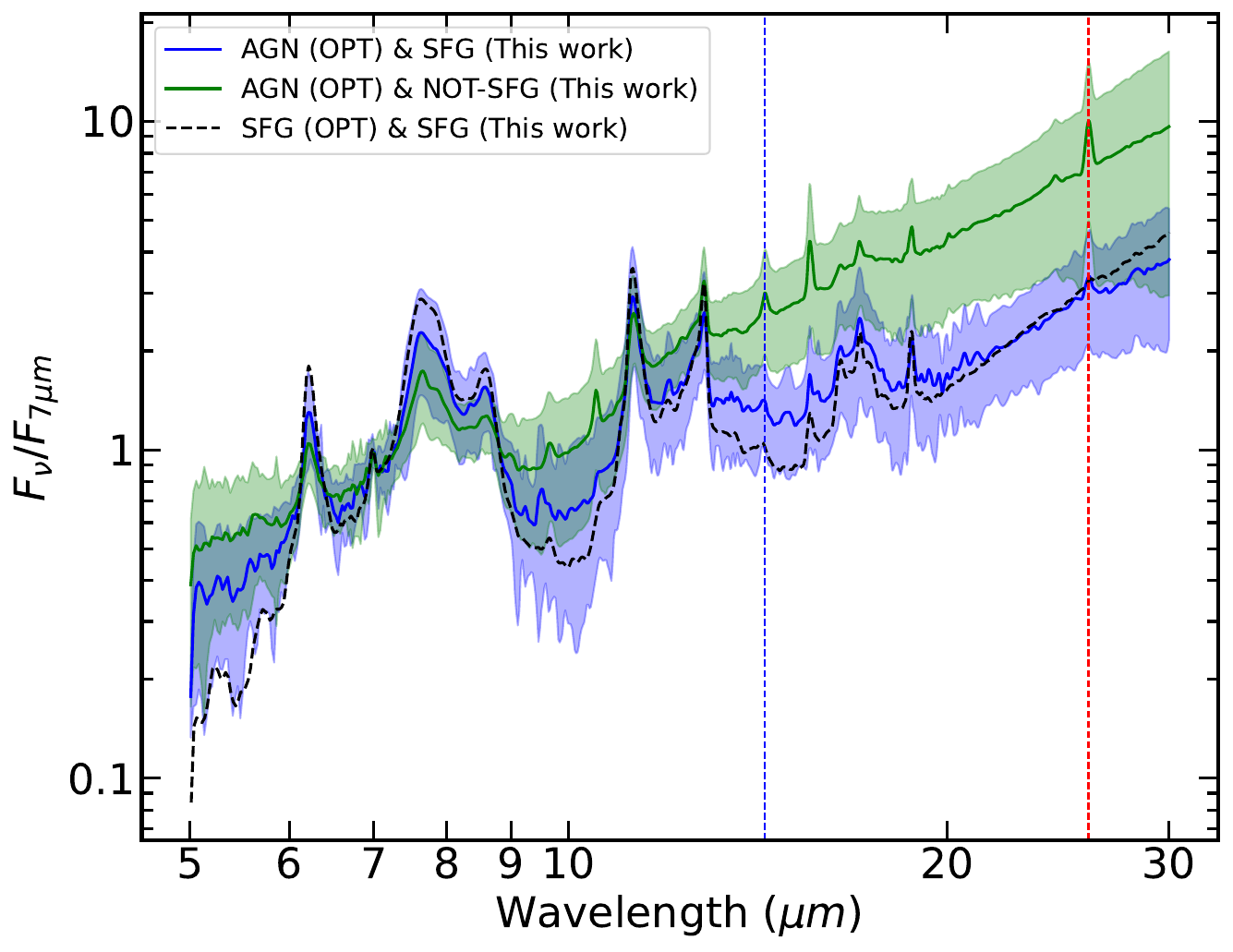}
\end{center}
\caption{Average mid-IR spectra of AGN galaxies (based on the BPT optical diagnostics) classified as star-forming (SFG) and non-star-forming (NOT-SFG) based on our diagnostic. The AGN galaxies predicted as star-forming galaxies (blue solid line) have similar average spectra with the normal star-forming galaxies (black dashed line). The AGN galaxies that have been predicted as not star-forming (blue solid line) have steeper spectra above 12\,$\mu$m than the BPT-AGN we predicted as star-forming galaxies. In addition, the average spectra of the AGN galaxies that we have identified as not star-forming show clear emission lines of [\ion{Ne}{V}] 14.3\,$\mu$m (blue dashed vertical line) and [\ion{O}{IV}] 25.9\,$\mu$m (red dashed vertical line) which are tell-tale signs of a strong radiation field source. The shaded area in both average spectra corresponds to the standard deviation of 1$\sigma$.}
\label{fig:mn_SED_AGN}
\end{figure}

In Fig. \ref{fig:mn_SED_AGN} we show these average spectra. The galaxies hosting AGN classified as star-forming have flatter spectra above 14\,$\mu$m than those classified as non star-forming. In addition, the latter show reduced PAH emission. In the same figure, the average spectra of the AGN misclassified as star-forming resembles that of a typical MS star-forming galaxy. Finally, the AGNs that have been predicted as non-star-forming galaxies, show prominent emission from forbidden lines with high ionization potential, such as [\ion{Ne}{V}] 14.3\,$\mu$m and [\ion{O}{IV}] 25.9\,$\mu$m which are generally absent from the AGN we classify as star forming.

We further analyze these optically classified galaxies using infrared AGN emission-line diagnostics. One such diagnostic employs the ratio of [\ion{O}{IV}] 25.9\,$\mu$m to the [\ion{Ne}{II}] 12.8\,$\mu$m as a measure of the fractional AGN contribution to the overall galaxy emission \citep{2002A&A...393..821S}. Similarly, the EW of the 6.2\,$\mu$m PAH feature serves as a proxy for the fractional starburst contribution to the galaxy's luminosity \citep{2007ApJ...656..148A}. Figure \ref{fig:Strum_Armus} shows the [\ion{O}{IV}]/[\ion{Ne}{II}] ratio as a function of 6.2\,$\mu$m PAH EW for the same sample of galaxies presented in Fig. \ref{fig:mn_SED_AGN} and \ref{fig:mn_SED_comp}. Red circles mark all galaxies that our diagnostic identifies as star-forming. All galaxies classified as star-forming exhibit low
AGN fractional contribution to their luminosity ($\lesssim $10\%), as determined by the [\ion{O}{IV}]/[\ion{Ne}{II}] ratio while they also show high starburst fractional contribution ($\sim$100\%), as indicated by the EW of the 6.2\,$\mu$m PAH feature. Furthermore, it is evident that all AGN and composite galaxies that have been misclassified as star-forming have lower AGN fractional contributions based on the [\ion{O}{IV}]/[\ion{Ne}{II}] ratio ($\lesssim $25\%) compared to the rest of the AGNs that were identified correctly as non star-forming, while at the same time showing prominent PAH emission based on the EW 6.2\,$\mu$m PAH feature showing significant starburst fractional contribution ($\gtrsim$50\%). This result ensures that our diagnostic is capable of identifying galaxies that their emission is dominated by star formation processes.

\begin{figure}[h]
  \resizebox{\hsize}{!}{\includegraphics{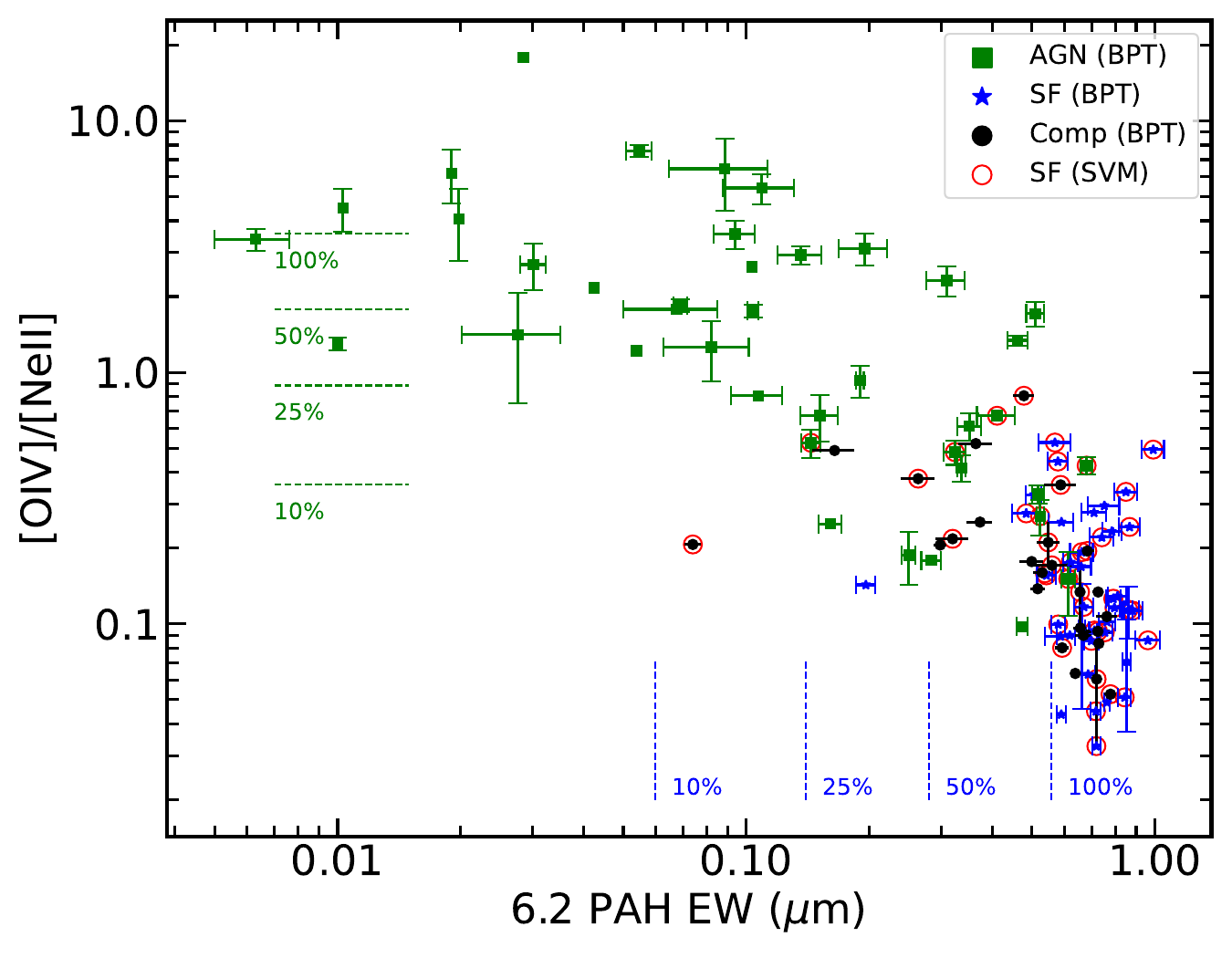}}
  \caption{The [\ion{O}{IV}] 25.9\,$\mu$m/[\ion{Ne}{II}] 12.8\,$\mu$m ratio is plotted against the EW of the 6.2\,$\mu$m PAH feature for optically identified star-forming (blue stars), composite (black disks), and AGN (green squares) galaxies. The green lines represent the average AGN fractional contribution to a galaxy’s emission, based on the [\ion{O}{IV}]/[\ion{Ne}{II}] ratio \citep{2002A&A...393..821S}. The blue lines denote the average starburst fractional contribution, determined from the EW of the 6.2\,$\mu$m feature \citep{2007ApJ...656..148A}. The red circle highlights any galaxy identified as star-forming based on our diagnostic tool. The uncertainties associated with each measurement are 1$\sigma$ error.}
  \label{fig:Strum_Armus}
\end{figure}

\subsection{Uncertainties and limitations of dust models}

The data used to train our diagnostic in this work are based on synthetic spectra constructed from the \cite{2007ApJ...663..866D} PAH emission templates, which represent a wide variety of dust conditions which however, are based on a simplified framework for modeling PAH features in galaxies. While this framework is applicable to typical star-forming galaxies, it may not be applicable to more extreme environments (e.g., starburst and/or low-metallicity galaxies) or in environments with complex activity (e.g., composite galaxies with a weak AGN contribution). These models assume a fixed set of PAH templates and parameterizations (e.g., grain size distribution, ionization fraction), and do not fully capture the environmental dependence (e.g., hardness of the starlight ionization field and PAH excitation in a self-consistently way) or intrinsic variability of PAH spectral shapes. Consequently, the learned representations may underperform in environments where PAH emission deviates significantly from the training set (e.g., low-metallicity systems). Future incorporation of more physically detailed models \citep[e.g.,][]{2021ApJ...917....3D} will likely improve both the fidelity and robustness of the predicted spectra. Nevertheless, given that our simulations are confined to MS star-forming galaxies, we anticipate that the measured PAH features from these simulations will not exhibit substantial deviations from spectra obtained from actual galaxies, as is demonstrated by the good agreement between the model SEDs and observations of typical star-forming galaxies \citep[e.g.,][]{2009A&A...507.1793N}.

\subsection{Redshift application range}

The use of broad-band features (i.e., photometry) can be robust to small changes in distances without requiring redshift corrections because the bands cover a broad area of the spectrum. Here, we investigate how the recall score of the star-forming galaxies would vary as a function of redshift. To accomplish this we apply our trained diagnostic on the simulated sample of galaxies (test set, see Sect. \ref{simgalspec}) by shifting their spectra to increasingly higher redshifts of $z=$ 0, 0.0125, 0.025, 0.05, 0.075, and 0.1. To evaluate the performance we keep the photometric bands fixed (as defined in Sect. \ref{secftrsid}) and we measure the corresponding flux of each feature in the redshifted spectrum. Figure \ref{fig:redshihft} presents the recall of star-forming galaxies as a function of redshift. There our recall is high for redshifts up to $z\sim0.02$ and falls linearly for $z>0.02$. We chose not to test the performance of our diagnostic beyond $z=0.1$ as the edge of our bluer custom band starts to leave the observing window of most infrared observatories.

\begin{figure}[h]
\begin{center}
\includegraphics[scale=0.39]{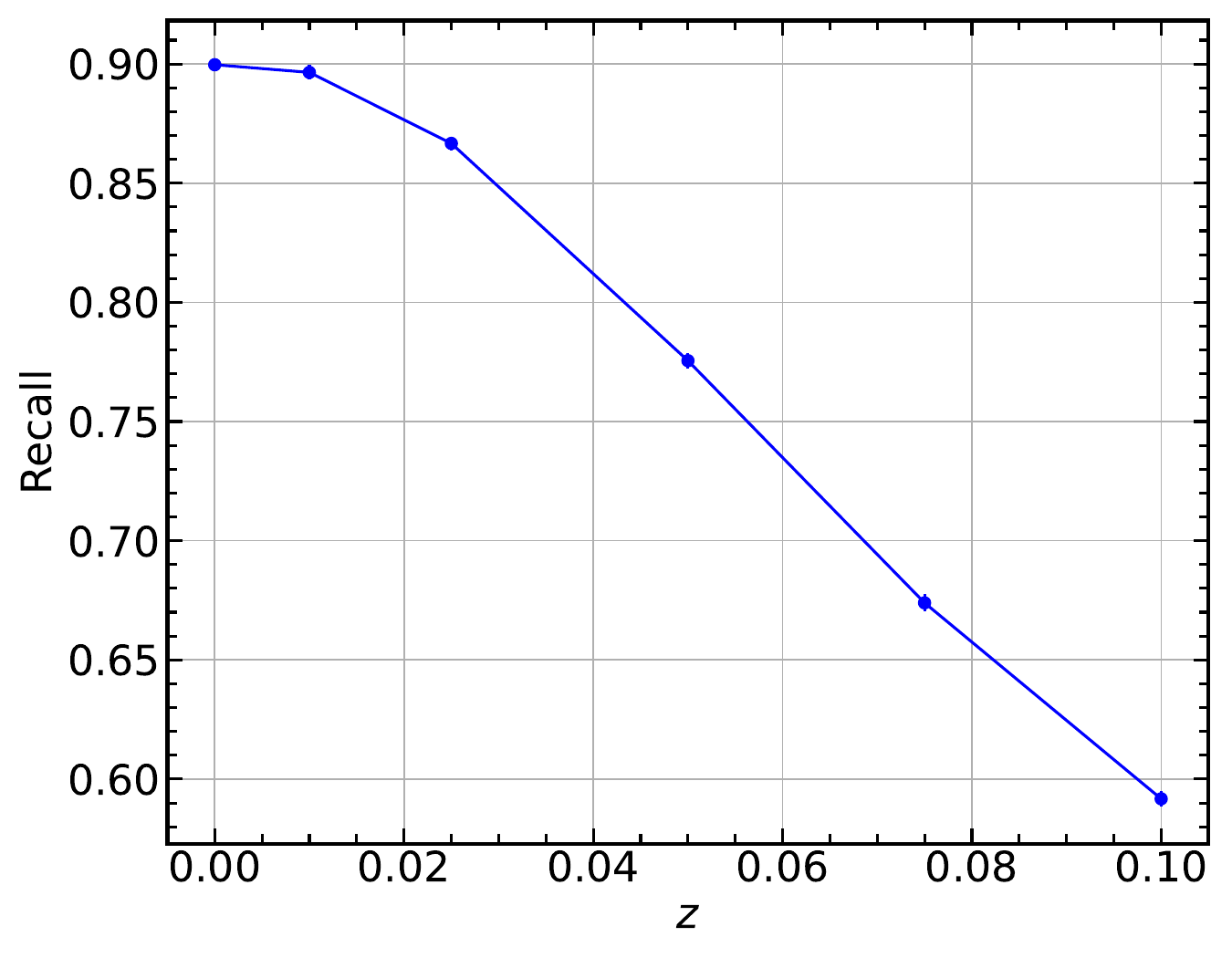}
\end{center}
\caption{The recall score of our diagnostic as a function of redshift. The performance was recorded by training the algorithm using rest-frame spectra and testing the performance on the test sample of the redshifted simulated spectra at $z=$ 0, 0.0125, 0.025, 0.05, 0.075, and 0.1. The scores are stable for up to $z\sim0.02$ and then they are dropping smoothly.}
\label{fig:redshihft}
\end{figure}

\begin{figure*}[!h] % Use figure* to span two columns
    \centering
    \begin{minipage}[t]{0.48\textwidth} % Left plot
        \centering
        \includegraphics[width=\textwidth]{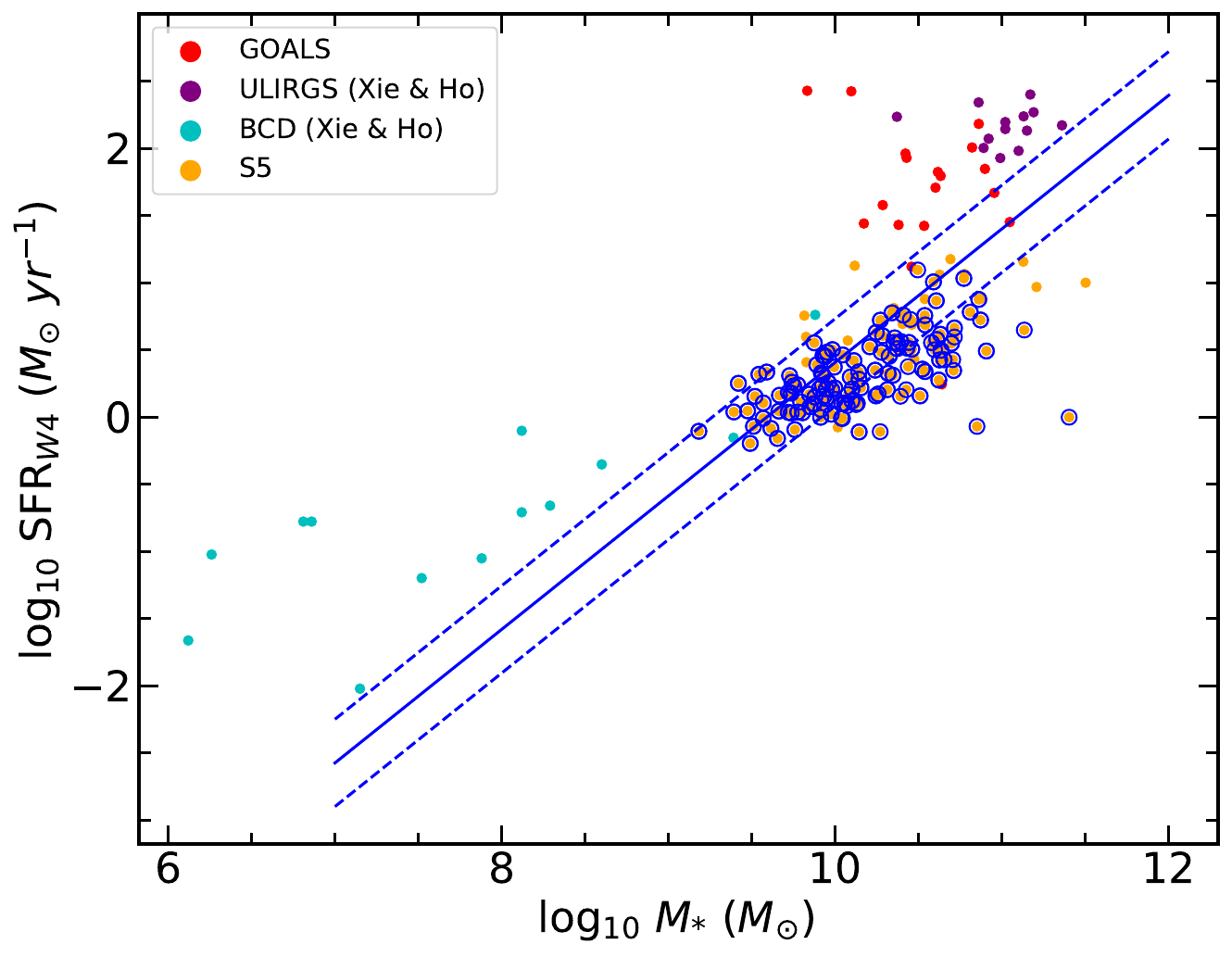} % Replace with your image file
    \end{minipage}
    \hfill % Adds horizontal space between the two minipages
    \begin{minipage}[t]{0.48\textwidth} % Right plot
        \centering
        \includegraphics[width=\textwidth]{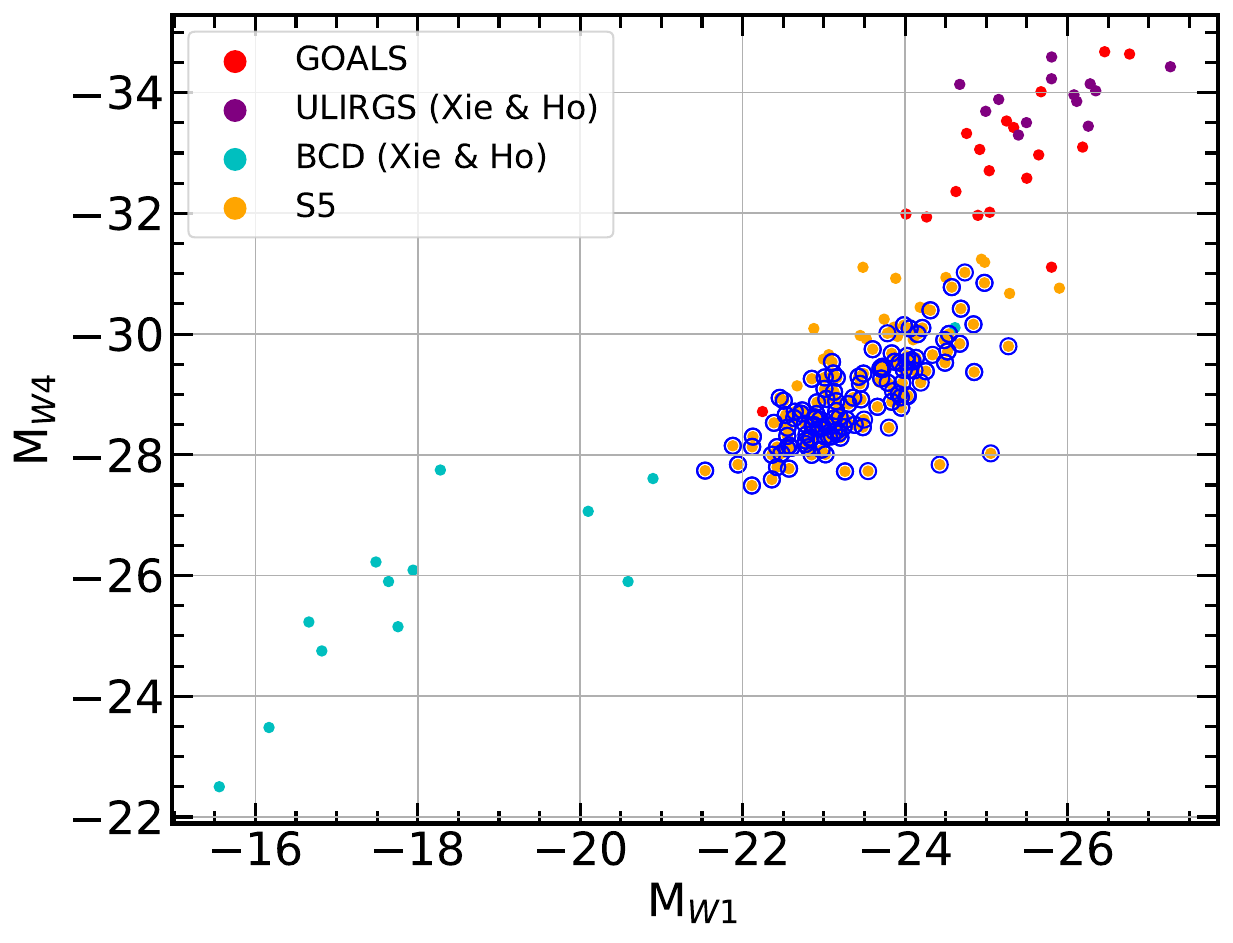} % Replace with your image file
    \end{minipage}
    \caption{Star formation rate against stellar mass for galaxies from several galaxy samples (left panel). We identify correctly almost all the galaxies that fall in the area on the MS star-forming galaxies. As expected, we miss all the galaxies from samples focused on BCDs and ULIRGs. The blue solid line shows the MS fit for star-forming galaxies from \cite{2011A&A...533A.119E} and the blue dashed lines are correspond to the $1\sigma$ error to this line. On the right panel we plot the absolute magnitude in band W4 against the absolute magnitude in band W1 for the same galaxies. The former serves as a proxy of the star formation rate while the latter as a proxy of the stellar mass of a galaxy. The star-forming galaxies that we have identified correctly (blue circles) are clustered on the main sequence while the star-forming galaxies that we miss tend to have higher M$_{W4}$ values or are extreme cases of galaxies (BCDs or ULIRGs).}
    \label{fig:MS_MW14} % Common label
\end{figure*}

\subsection{Applicability limits}

Our diagnostic tool was developed to identify MS star-forming galaxies, which are distinct from other types of galaxies that host star formation, such as starburst galaxies ULIRGs and BCDs. The MS galaxies have different infrared properties (e.g., colors) when compared to other extreme classes of star-forming galaxies such as BCDs or ULIRGs as the latter two both have red IR colors \citep{2016ApJ...832..119H}. In this section, we aim to assess the applicability of our diagnostic on different samples of galaxies as defined by their IR properties (e.g., colors and magnitudes) by identifying areas on color-color and magnitude-magnitude diagrams. 

We seek the locus within the MS where our diagnostic correctly identifies star-forming  galaxies, in order to identify the limits within which it works optimally. For that reason, the samples of GOALS, ULIRGs, BCDs, and S5 are cross-matched with the WISE All-sky catalog to acquire their W4 magnitudes, and utilizing the SFR-W4 SFR calibration of \cite{2017ApJ...850...68C} we calculate the SFRs of all galaxies. Left panel of Fig. \ref{fig:MS_MW14} shows our results from the application of our diagnostic to the aforementioned samples of galaxies on the SFR-stellar mass plot. We adopt the MS definition from \cite{2011A&A...533A.119E}. In that figure, BCDs and ULIRGs deviate from the best-fit line that contains normal SF galaxies.

% \begin{figure}[h]
% \begin{center}
% \includegraphics[scale=0.39]{plots/SRF_M_star_MS.pdf}
% \end{center}
% \caption{Star-formation rate (log SFR) against stellar mass (log $M_{*}$) for galaxies from several galaxies surveys. We identify correctly almost all the galaxies that fall in the area on the MS star-forming galaxies. As expected, we miss all the galaxies from surveys focused on BCDs and ULIRGs. The blue solid line is the MS fit for star-forming galaxies from \cite{2011A&A...533A.119E} and the blue dashed lines are correspond to the $1\sigma$ error to this line.}
% \label{fig:MS}
% \end{figure}

However, stellar mass and SFR estimations can sometimes be difficult to acquire. For this reason, we also consider infrared photometric bands from all-sky surveys as these will include a large number of galaxies making it easily applicable for large datasets. Thus, we used the four WISE bands (W1, W2, W3, and W4) to calculate the absolute magnitudes of W1 and W4. In Fig. \ref{fig:MS_MW14} (right panel) we plot the W4 absolute magnitude against the W1 absolute magnitude. The former provides an approximation of the star-formation rate of a star-forming galaxy \citep{2017ApJ...850...68C} while the latter provides an estimation of the stellar mass. 
% There we see that both BCDs and ULIRGs have higher luminosities in W4.
% Comparing the two panels of \ref{} we see that our diagnostic can be applied reliably for 9.5 < log $M_{*}$ < 11 corresponding to -25 < M$_{W1}$ < -22 and 0 < log SFR < 1 corresponding to -27.5 < log M$_{W4}$ < -31.

% \begin{figure}[h]
% \begin{center}
% \includegraphics[scale=0.42]{plots/SRF_M_star_WISEabs.pdf}
% \end{center}
% \caption{Plot of the absolute magnitude in band W4 against the absolute magnitude in band W1 for the samples of galaxies introduced in Fig. \ref{fig:MS}. The former serves as a proxy of the star formation rate while the latter as a proxy of the stellar mass of a galaxy. The star-forming galaxies that we have identified correctly are clustered on the main sequence while the star-forming galaxies that we miss tend to have higher M$_{W4}$ values or are extreme cases of galaxies (BCDs or ULIRGs).}
% \label{fig:abs_w4_w1}
% \end{figure}

\begin{figure}[h]
\begin{center}
\includegraphics[scale=0.43]{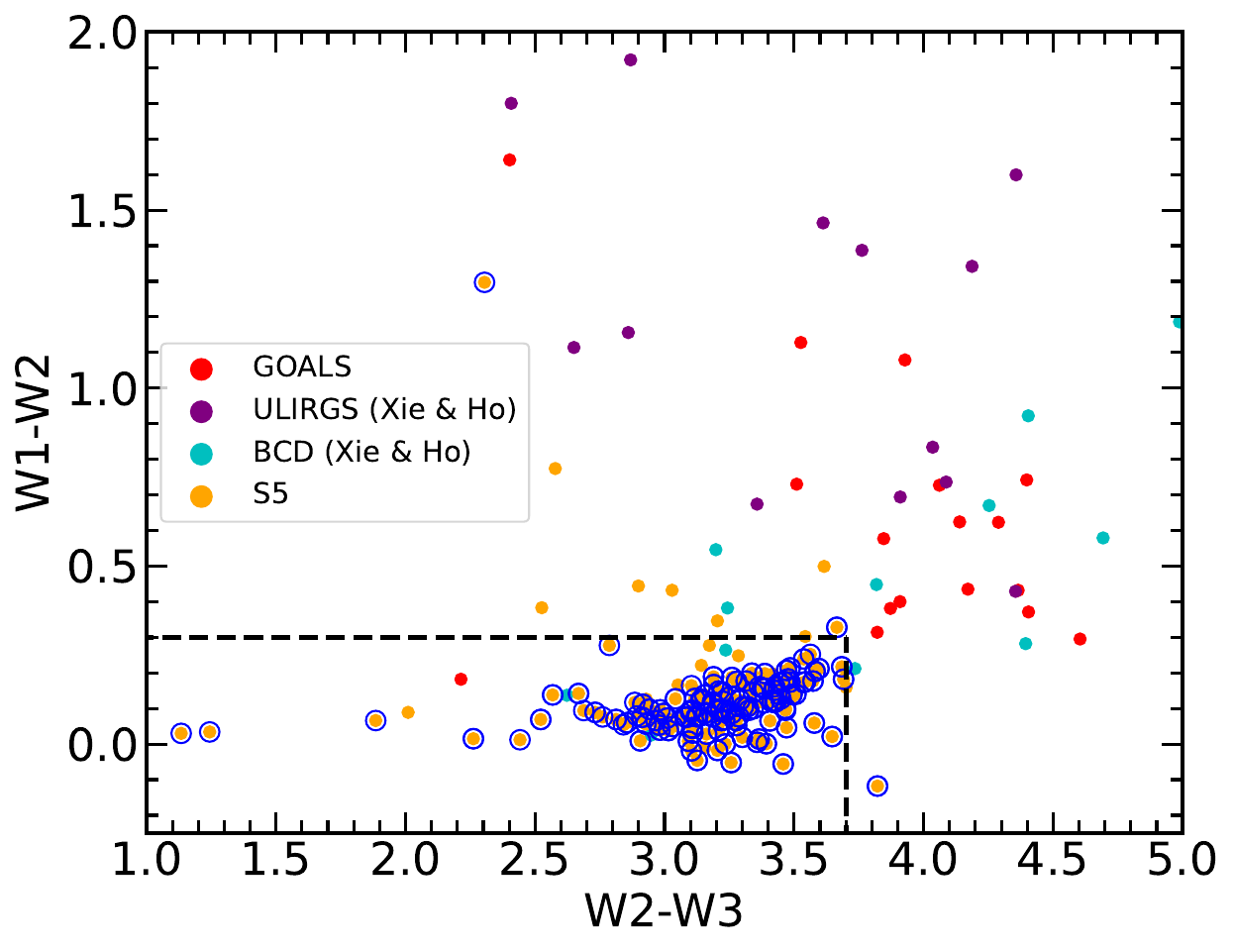}
\end{center}
\caption{Color-color plot of $W1-W2$ against the $W2-W3$ for the same samples of galaxies introduced in Fig. \ref{fig:MS_MW14}. The normal star-forming galaxies we have identified correctly occupy the lower area of this plot while extreme cases of galaxies that may be related with star formation (BCDs and ULIRGs) are scattered having redder $W1-W2$ and $W2-W3$ colors.The black dashed lines delineate the optimal operational area of our diagnostic system.}
\label{fig:cc_w1w2w2_real_gal_bndrs}
\end{figure}

Furthermore, a widely used IR color-color plot is the $W1-W2$ versus $W2-W3$. in Fig. \ref{fig:cc_w1w2w2_real_gal_bndrs}, we present the of $W1-W2$ versus $W2-W3$ color-color plot, utilizing the same sample of galaxies as in in Fig. \ref{fig:MS_MW14}. In Fig. \ref{fig:cc_w1w2w2_real_gal_bndrs}, we see that MS star-forming galaxies occupy a specific region on this color-color plot, characterized by bluer $W1-W2$ colors than the the more extreme star-forming galaxies (BCDs and ULIRGs). This region is bounded by two straight lines of $W1-W2$ < 0.3 and $W2-W3$ < 3.7. Based on this observation we propose this selection criteria as a test for the applicability of our diagnostic. Additionally, in the same figure, we observe that the MS star-forming galaxies which our diagnostic failed to identify correctly (9.1\% of the MS star-forming galaxies from the S5 sample) have redder $W1-W2$ and $W2-W3$ colors compared to those correctly identified. Galaxies exhibiting such IR colors can be AGN hosts or extreme starbursts \citep{2016ApJ...832..119H,2023A&A...679A..76D}.

\subsection{Comparison with other methods}

The custom diagnostic bands we defined for this diagnostic (Sect. \ref{secftrsid}) have the advantage of targeting specific spectral features, offering high discriminating power. They also provide unbiased measurements, unlike existing IR diagnostics (e.g., emission lines and 6.2\,$\mu$m EW) that require continuum subtraction and assumptions about the underlying dust continuum, which can lead to biased EWs by a factor of $\sim$2-3 \citep{2007ApJ...656..770S}.

\begin{table}
\centering
\caption{Comparison of the activity classification between our diagnostic and the 6.2\,$\mu$m PAH EW classifier (EW6.2).}
\begin{tabular}{ccccccc}
\multicolumn{1}{l}{}               & 
\multicolumn{4}{c}{BPT-SFG (138)}                               \\ \hline
\multicolumn{3}{r}{EW6.2}  \\
\multirow{10}{*}{\rotcell{This~work}} %
& & SFG & AGN & Comp & Total  \\ \hline
 & SFG   & 124 & 0 & 0 & 124 \\
 & NON-SFG   & 14 & 0 & 0  & 14 \\
 & Total & 138 & 0 & 0 &  &  
\\ \\ \\ 
\multicolumn{6}{c}{BPT-AGN (127)}                           \\ \hline
\multicolumn{3}{r}{EW6.2}  \\
\multirow{10}{*}{\rotcell{This~work}}
& & SFG & AGN & Comp & Total  \\ \hline
% & & SFG & AGN & LINER  
%  & Total  \\ \hline
 & SFG     & 14 & 0 & 2 & 16 \\
 & NON-SFG & 30 & 51 & 30  & 111 \\
 & Total  & 44 & 51 &  32 &  
 \\ \\ \\
\multicolumn{6}{c}{BPT-composite (108)}                               \\ \hline
\multicolumn{3}{r}{EW6.2}  \\
\multirow{10}{*}{\rotcell{This~work}}
& & SFG & AGN & Comp & Total  \\ \hline
%  & Total  \\ \hline
 & SFG     & 31 & 1 & 1 & 33 \\
 & NON-SFG & 62 & 8 & 5  & 75 \\
 & Total   & 93 & 9 & 6 & \\
 \\ \\
\end{tabular}
\label{tab:wha}
\tablefoot{The SFG, AGN, and composite galaxies are from the S5 sample (Sect. \ref{obs_spec}). Here we compare our diagnostic with the classification method of \cite{2007ApJ...656..148A} (EW6.2), against the optical spectroscopic method (BPT) which we consider as ground truth.}
\end{table}

In order to compare the performance of our diagnostic with other similar diagnostic methods, we used the diagnostic based on EW of 6.2\,$\mu$m PAH introduced in Sect. \ref{s51}. Galaxies with an EW below 0.2\,$\mu$m are classified as AGN while those with an EW between 0.2 and 0.5\,$\mu$m are classified as composite, and those with an EW above 0.5\,$\mu$m are designated as star-forming galaxies.

Our comparison is based on star-forming galaxies from the S5 sample because they represent a wide range of MS galaxy conditions (see Sect. \ref{obs_spec}). However, this sample lacks galaxies of other activity types, such as composite and AGN galaxies. To address this limitation, we supplemented our sample with composite and AGN galaxies from the IDEOS sample (Sect. \ref{obs_spec}). 

Although the IDEOS sample provides various observables for its galaxies, it does not offer activity classifications based on optical methods. Therefore, we cross-matched the IDEOS catalog with the SDSS DR8 to obtain optical emission line fluxes, such as H$\alpha$, H$\beta$, [\ion{N}{II}], [\ion{O}{III}], and [\ion{S}{II}] and determined their activity based on the emission-line ratios of log$_{10}$([\ion{N}{II}]/H$\alpha$), log$_{10}$([\ion{S}{II}]/H$\alpha$), and log$_{10}$([\ion{O}{III}]/H$\beta$). To maintain consistency with the real galaxy sample (see Sect. \ref{obs_spec}), these galaxies have to be classified using an optical emission line diagnostic. Thus, we employed the classification tool developed by \cite{2019MNRAS.485.1085S}, which is based on the location of each galaxy in the three-dimensional emission-line ratio space using the aforementioned spectral lines. This approach is advantageous to the traditional two-dimensional diagnostic diagrams as it simultaneously considers three of the strongest optical emission lines avoiding contradictory classifications between the different two-dimensional diagnostics.

To ensure that high quality data are used in this comparison we filtered the S4 and IDEOS spectra by selecting those with S/N greater than 5 at the 14 $\mu$m continuum. In addition we considered, only measurements with S/N greater than 3 for the EW of 6.2 $\mu$m PAH feature. For the quality of the optical classifications, we selected galaxies with an S/N greater than 5 in all the emission lines used (i.e., H$\alpha$, H$\beta$, [\ion{O}{III}], [\ion{N}{II}], and [\ion{S}{II}]). Our final comparison sample consists of 138 star-forming galaxies, 127 AGN, and 108 composite galaxies. We compare our method with the EW6.2 diagnostic and the optical classification, which serves as the ground truth. The results of this comparison are summarized in Table \ref{tab:wha}.

From these classification results, our diagnostic correctly identifies 90\% of the star-forming galaxies, while the EW6.2 method correctly identifies all of them. Closer inspection of the mid-infrared SEDs of these misclassified galaxies reveals that they do not share the same properties as typical MS star-forming galaxies since their mid-IR SEDs are similar to the composite galaxies that our diagnostic predicted as as not star-forming galaxies shown in Fig. \ref{fig:mn_SED_comp}.

For the optically classified AGN galaxies, our diagnostic misclassified only one AGN as a star-forming galaxy, whereas the EW6.2 method misclassified nearly half of them as star-forming. Upon examining the SEDs, we found that the one galaxy our diagnostic misclassified as star-forming has an SED similar to a typical MS star-forming galaxy (see Fig. \ref{fig:mn_SED_AGN}).

When comparing the activity class of composite galaxies, our diagnostic results in classifying 33\% of them as star-forming and the other 67\% as non-star-forming. In contrast, the EW6.2 method identifies almost all composite galaxies as star-forming, with only one classified as composite and another as AGN. By visually inspecting the SEDs of the composite galaxies classified by our diagnostic, we found that those predicted as star-forming have SEDs consistent with the average spectrum of a typical star-forming galaxy (e.g., they lack PAH emission, they have steeper IR continuum), while those predicted as non-star-forming do not exhibit similarities with star-forming galaxies (see Fig. \ref{fig:mn_SED_comp} and Sect. \ref{comp_av_spec} for a more detailed discussion). This result shows that our diagnostic can identify galaxies that, despite exhibiting high 6.2\,$\mu$m EW which is usually linked to star-forming galaxies, they cannot be characterized as main sequence star-forming galaxies. The presence of the steeper slope indicates more intense ionization fields illuminating the dust than those typically found in normal MS star-forming galaxies.

In conclusion, we observe that the main differentiating factor in all these cases is the continuum emission at 22\,$\mu$m. This indicates that our diagnostic is sensitive to the stronger dust emission that can be a result of a harder radiation field (i.e., higher temperature dust), even in the presence of significant PAH emission, which in some cases seems to confuse the EW6.2 diagnostic.

\subsection{Potential application to JWST/MIRI spectra}

Since PAH features are broad spectral features present over a wide spectral range we expect that our diagnostic will be applicable to spectra acquired with different instruments regardless of their resolving power. JWST in particular, enables the acquisition of high quality spectra from a much larger population of galaxies than was possible with previous instruments.

In order to verify the applicability of our diagnostic to data obtained with JWST, we applied it to data obtained with the MIRI spectrograph onboard the JWST. We used spectra extracted from galaxy cores that are known to host AGN (NGC 7469) and galaxies that have star forming regions scattered across the disk (VV114). The extraction of the spectra was performed using the Continuum And Feature Extraction ({CAFE; \cite{2025ascl.soft01001D}}\footnote[4]{\url{https://github.com/GOALS-survey/CAFE}}). Then, we measured the intensity of our four diagnostic bands on their rest-frame spectra. We found that all the AGN spectra are correctly identified as not being MS star-forming galaxies while the spectra extracted from the \ion{H}{II} regions are correctly identified as resembling MS star-forming galaxies. This exercise proves that our diagnostic can indeed applied to spectra recorded with different spectral resolution.

\subsection{A Photometric diagnostic optimized for JWST}

As discussed in Sect. \ref{secftrsid}, our custom photometric bands are designed to specifically target specific, physically motivated, PAH and continuum features that have demonstrated exceptional effectiveness in identifying MS star-forming galaxies from their IR spectra. However, photometric observations are generally easier to obtain than spectroscopy, and with JWST’s growing archival database, it is both timely and valuable to develop a diagnostic tailored in identifying MS star-forming galaxies using JWST photometry alone. While previous missions such as WISE and Spitzer (e.g., IRAC and MIPS) have offered broad mid-infrared coverage, their wider filter profiles often blend PAH emission features with the underlying continuum, which can dilute the PAH characteristics we aim to isolate. In contrast, JWST’s narrower, more precisely defined filters provide an opportunity to isolate key PAH signatures with greater clarity, improving the reliability of photometric diagnostics. Thus, prompted by the results presented in the previous sections we define an additional diagnostic tool that is tailored to the JWST photometry.

To accomplish this objective, we identify JWST bands that span the 5-24\,$\mu$m wavelength range form the MIRI instrument. From the diverse available filters, we select those that resample the wavelength ranges more closely to our photometric band scheme (Table \ref{pah_bands}). This way we select the F560W, F770W, F1130W, and F2100W filters. Based on these bands we calculate the flux ratios of F560W/F2100W, F770W/F2100W, and F1130W/F2100W as discriminating features equivalent to our diagnostic scheme.

Afterwards, we train and evaluate our diagnostic according to the procedures described in Sect. \ref{trlg} and Sect. \ref{sec_eval_prf} respectively. More specifically, we started by calculating the synthetic fluxes on the simulated spectra (Sect. \ref{simgalspec}) by convolving them with the response of each selected JWST filter (filter response curves taken from {Spanish Virtual Observatory (SVO)}\footnote[5]{\url{https://svo.cab.inta-csic.es/main/index.php}}). After training the new diagnostic, we evaluated its performance on the sample of real galaxies (Sect. \ref{obs_spec} and Table \ref{tab:wha}). This sample does not have available JWST photometry. For this reason, we calculated their synthetic fluxes by convolving their Spitzer/IRS spectra as we did for the simulated spectra. In Table \ref{conf_mat_JWST} we present the results of the application of the JWST-optimized diagnostic on the JWST photometry calculated for the real galaxy sample presented in Table \ref{tab:wha}.

\begin{table}[ht]
\centering
\caption{Classification results for a sample of real galaxies based on the diagnostic tailored for JWST/MIRI photometry.}
\begin{tabular}{lccccc}
       &         & SFG & AGN & Composite & Total \\
       &         & (138)  & (127)  & (108) & (373)    \\ \hline
\multirow{3}{*}{\rotatebox[origin=c]{90}{\shortstack{JWST\\diagnostic}}} 
       & SFG     & 131 & 35  & 98 & 264      \\
       & NON-SFG & 7   & 92  & 10   & 109       \\
       & Total   & 138 & 127 & 108 &       \\
\end{tabular}
\label{conf_mat_JWST}
\end{table}

Based on the results presented in Table \ref{conf_mat_JWST}, we see several notable distinctions between the JWST-optimized tool and the tool described in the previous sections. First, the new diagnostic shows improved performance in identifying MS star-forming galaxies, increasing from 90\% to 95\%. However, this improvement comes with trade-offs. Of the 127 AGNs in the sample, 35 (28\%) are misclassified as MS star-forming galaxies-doubling the AGN contamination rate compared to the previous diagnostic (which was 12.5\%). Additionally, the JWST/MIRI filter-based diagnostic classifies the majority of composite galaxies ($\sim$91\%) as MS star-forming, contributing further contamination. Figure \ref{fig:mn_SED_comp} shows that a substantial subset of our sample of composites ($\sim$67\%) exhibit prominent AGN signatures, such as [\ion{Ne}{V}] and [\ion{O}{IV}] emission lines in their spectra, suggesting that they are not MS star-forming galaxies.

These results indicate that while the JWST diagnostic slightly improves the identification of MS star-forming galaxies, it also increases misclassification of AGNs and composites, limiting its reliability for defining a clean MS star-forming sample. This reinforces the conclusion that our custom-selected bands offer a more optimal feature set for discrimination. The reason why the JWST filters underperform in this context may be that, although the first three filters span similar wavelength range as our scheme, the F2100W filter spans a wider range (17.9–24.5\,$\mu$m). This prevents it from isolating the warm dust continuum effectively, as it overlaps with PAH features in the 15-20\,$\mu$m region. Nonetheless, this diagnostic can be used to obtain a baseline sample of MS galaxies from JWST photometric surveys, albeit with the aforementioned limitations.

\section{Conclusions} \label{concl}

In this work, we developed a diagnostic for MS star-forming galaxies in the local Universe by combining a library of simulated IR spectra and machine learning methods. Our results can be summarized as follows.
\begin{enumerate}
  \item We build a library of synthetic UV to IR spectra of non-AGN galaxies, by sampling parameters for the stellar populations and ISM obtained from extensive observational studies. We find that the optical and near/mid-IR colors of these simulated spectra are consistent with the observed colors of large galaxy samples. Nevertheless, we observe a general agreement between the simulated galaxies’ UV colors and the actual sample. Notably, objects located in the middle of the $NUV-r$ band exhibit a systematic offset.
  \item We build a diagnostic tool based on photometric bands centered on dust emission features (PAHs and continuum). This diagnostic can reliably identify MS sequence star-forming galaxies with minimal contamination by AGN. The use of broadband photometry circumvents the process of estimating and subtracting the underlying continuum, as such a process can introduce inherent biases.
  \item The utilization of broad bands and the fact that PAHs and continuum features employed here are effectively insensitive to the resolution of the spectrograph utilized to record the spectra, our diagnostic can be applied to spectra acquired from other observatories without any modifications, such as the JWST.
  \item In addition to our diagnostic, we provide a separate diagnostic tool optimized for JWST photometry using only four JWST/MIRI filters: F560W, F770W, F1130W, and F2100W. While the JWST diagnostic exhibit reduced performance compared to our original scheme, it remains highly relevant for future analyses based on JWST data as more data becomes available.
 \end{enumerate}
Extensions of this diagnostic include a study of the sensitivity of this diagnostic to the presence of weak AGN, and the characterization of extreme starbursts (e.g. BCDs and ULIRGs).

\section{Data availability}

Our diagnostic, including the version tailored for JWST/MIRI photometry, will be made publicly available on GitHub\footnote[6]{\url{https://github.com/BabisDaoutis/MSMIRdiagnostic}}, accompanied by detailed documentation and usage instructions.

\begin{acknowledgements}
We would like to express our sincere gratitude to the anonymous referee for their thoughtful comments and constructive suggestions, which have helped improve the clarity and quality of the manuscript. We would like to express our sincere gratitude to Dr. Angelos Nersessian and Dr. Emmanuel Xilouris for results from the dustPedia analysis that are used in our analysis, Dr. Benjamin Johnson for insightful discussions on the SED simulations. C.D. acknowledges support from the Public Investments Program through a Matching Funds grant to the IA-FORTH. The research leading to these results has received funding from the European Research Council under the European Union’s Seventh Framework Programme (FP/2007-2013)/ERC Grant Agreement n. 617001, the European Union’s Horizon 2020 research and innovation programme under the Marie Skłodowska-Curie RISE action, Grant Agreement n.873089 (ASTROSTAT-II), and the Smithsonian Astrophysical Observatory Predoctoral Program with funding under the NASA grant 80NSSC21K0078. This research has made use of the NASA/IPAC Infrared Science Archive, which is funded by the National Aeronautics and Space Administration and operated by the California Institute of Technology. DustPedia is a collaborative focused research project supported by the European Union under the Seventh Framework Programme (2007-2013) call (proposal no. 606847). The participating institutions are: Cardiff University, UK; National Observatory of Athens, Greece; Ghent University, Belgium; Université Paris Sud, France; National Institute for Astrophysics, Italy and CEA, France. Funding for the Sloan Digital Sky 
Survey IV has been provided by the 
Alfred P. Sloan Foundation, the U.S. 
Department of Energy Office of 
Science, and the Participating 
Institutions. 

SDSS-IV acknowledges support and 
resources from the Center for High 
Performance Computing  at the 
University of Utah. The SDSS 
website is www.sdss.org.

SDSS-IV is managed by the 
Astrophysical Research Consortium 
for the Participating Institutions 
of the SDSS Collaboration including 
the Brazilian Participation Group, 
the Carnegie Institution for Science, 
Carnegie Mellon University, Center for 
Astrophysics | Harvard \& 
Smithsonian, the Chilean Participation 
Group, the French Participation Group, 
Instituto de Astrof\'isica de 
Canarias, The Johns Hopkins 
University, Kavli Institute for the 
Physics and Mathematics of the 
Universe (IPMU) / University of 
Tokyo, the Korean Participation Group, 
Lawrence Berkeley National Laboratory, 
Leibniz Institut f\"ur Astrophysik 
Potsdam (AIP),  Max-Planck-Institut 
f\"ur Astronomie (MPIA Heidelberg), 
Max-Planck-Institut f\"ur 
Astrophysik (MPA Garching), 
Max-Planck-Institut f\"ur 
Extraterrestrische Physik (MPE), 
National Astronomical Observatories of 
China, New Mexico State University, 
New York University, University of 
Notre Dame, Observat\'ario 
Nacional / MCTI, The Ohio State 
University, Pennsylvania State 
University, Shanghai 
Astronomical Observatory, United 
Kingdom Participation Group, 
Universidad Nacional Aut\'onoma 
de M\'exico, University of Arizona, 
University of Colorado Boulder, 
University of Oxford, University of 
Portsmouth, University of Utah, 
University of Virginia, University 
of Washington, University of 
Wisconsin, Vanderbilt University, 
and Yale University.

\end{acknowledgements}


% \begin{thebibliography}{}

%   \bibitem[Baker(1966)]{baker} Baker, N. 1966,
%       in Stellar Evolution,
%       ed.\ R. F. Stein,\& A. G. W. Cameron
%       (Plenum, New York) 333

%   \bibitem[Balluch(1988)]{balluch} Balluch, M. 1988,
%       A\&A, 200, 58

%   \bibitem[Cox(1980)]{cox} Cox, J. P. 1980,
%       Theory of Stellar Pulsation
%       (Princeton University Press, Princeton) 165

%   \bibitem[Cox(1969)]{cox69} Cox, A. N.,\& Stewart, J. N. 1969,
%       Academia Nauk, Scientific Information 15, 1

%   \bibitem[Mizuno(1980)]{mizuno} Mizuno H. 1980,
%       Prog. Theor. Phys., 64, 544
   
%   \bibitem[Tscharnuter(1987)]{tscharnuter} Tscharnuter W. M. 1987,
%       A\&A, 188, 55
  
%   \bibitem[Terlevich(1992)]{terlevich} Terlevich, R. 1992, in ASP Conf. Ser. 31, 
%       Relationships between Active Galactic Nuclei and Starburst Galaxies, 
%       ed. A. V. Filippenko, 13

%   \bibitem[Yorke(1980a)]{yorke80a} Yorke, H. W. 1980a,
%       A\&A, 86, 286

%   \bibitem[Zheng(1997)]{zheng} Zheng, W., Davidsen, A. F., Tytler, D. \& Kriss, G. A.
%       1997, preprint
% \end{thebibliography}

\bibliographystyle{aa}
\bibliography{references}

\begin{appendix} 

\section{Optimization of significant parameters} \label{apndx1}

As machine learning algorithms often have a lot of parameters that take a could potentially take lot of values. This makes them highly versatile offering them the ability to adapt to a huge number of datasets. Thus, a crucial step towards defining a new diagnostic tool is to optimize it so that the algorithm could tailored to fit the individual needs of a problem making it work at its peak performance. 

Optimization typically involves selecting parameters that significantly impact the algorithm's performance. For each parameter, a range of values is chosen, and a grid search explores all possible combinations. This process utilizes cross-validation, a robust technique for evaluating and improving the performance of machine learning models by reducing overfitting and ensuring generalizability.

In k-fold cross-validation, the dataset is divided into k equally sized folds. The model is trained on k-1 folds and tested on the remaining fold, with this process repeated k times to ensure each fold serves as the test set once. The results are averaged to provide a comprehensive performance metric. This method offers a better performance estimation compared to a single train-test split and makes full use of the available data, as each data point is included in both training and test sets across iterations.

\begin{figure}[h]
\begin{center}
\includegraphics[scale=0.39]{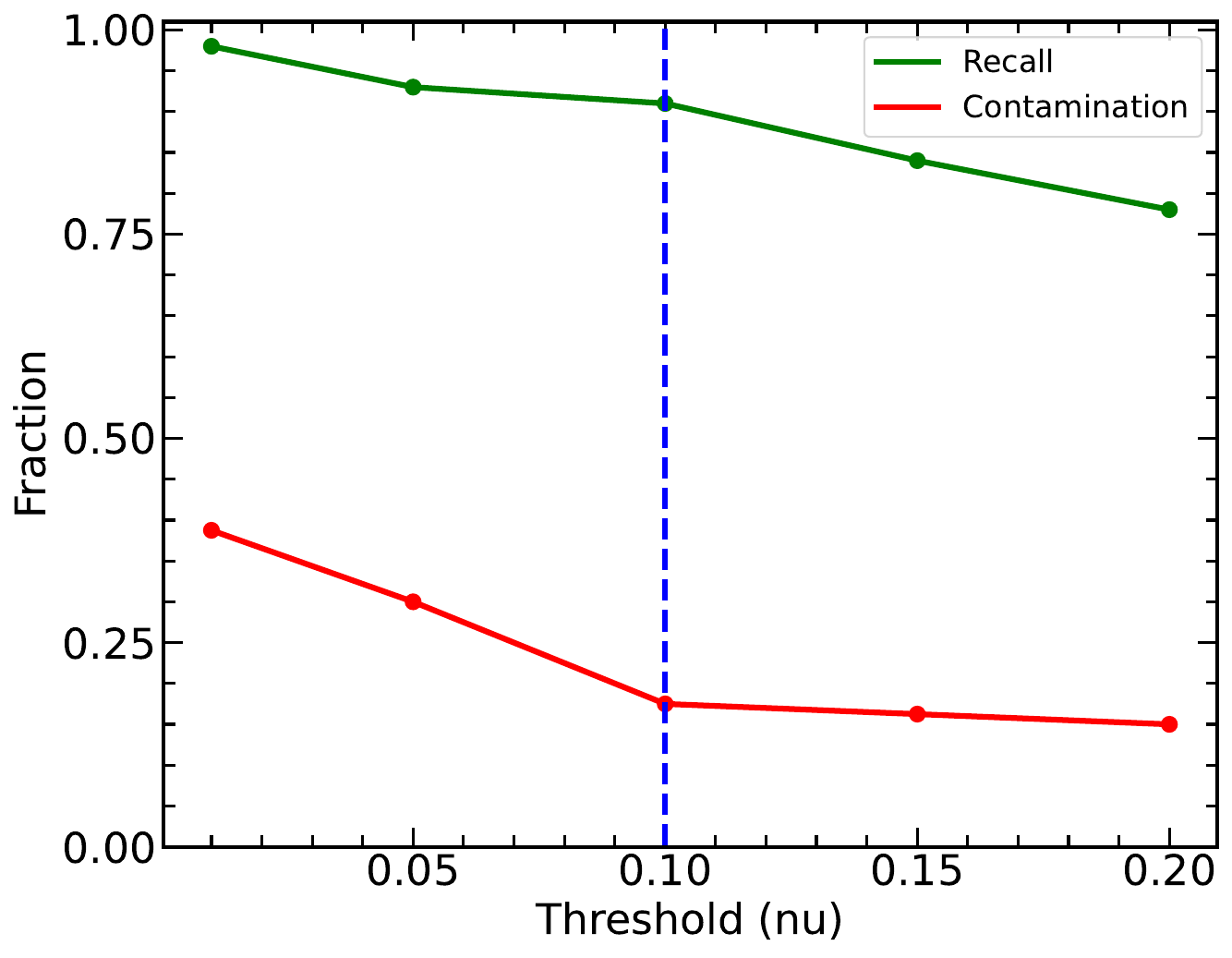}
\end{center}
\caption{Plot showing the recall for star-forming galaxies, indicated by the green line, and contamination from AGN, indicated by the red line. Both metrics were calculated on the observed sample (Sect. \ref{obs_spec}) as a function of the \texttt{oneclassSVM} nu parameter (i.e., population selection threshold). We observe that contamination starts to increase rapidly after \texttt{nu} = 0.9, marked by a blue horizontal line, which designates our selected value for \texttt{nu}.}
\label{fig:opitmization}
\end{figure}

Here, we are dealing with an algorithm designed to recognize only one population of objects. During its training, the \texttt{oneclassSVM} fits a surface in the selected feature space to encompass a user-defined fraction of the whole population. Data inside these boundaries are considered normal cases, while everything outside is considered an outlier. Thus, an assumption is made about what fraction of the given data represents the true population.

We perform the optimization of our diagnostic in two steps. The first step involves identifying the parameters that significantly impact performance. This step has two main purposes: to prevent the algorithm from overfitting the data and to achieve the best performance. After investigation, we found that the parameters with the highest impact on performance are \texttt{gamma} and \texttt{kernel}. Using exclusively the mock sample of galaxies (see Sect. \ref{simgalspec}) and the k-fold cross-validation method, we determined the best values for these parameters, as shown in Table \ref{oneclassSVM_params}. The remaining parameters are kept at their default values as specified by \texttt{oneclassSVM}.

\begin{table}[ht]
\caption{Parameters of the one-class SVM, their ranges, and best values.}
\centering
\begin{tabular}{l c c}
\hline\hline
Parameter & Range & Best Value \\
\hline
\texttt{gamma} & ‘scale’, ‘auto’ & 'scale' \\
\texttt{kernel} & 'linear', 'poly', 'rbf', 'sigmoid' & 'linear' \\
\texttt{nu} & 0.01 ,0.05, 0.10, 0.15, 0.2 & 0.1\\
\hline
\end{tabular}
\label{oneclassSVM_params}
\tablefoot{The \texttt{gamma} and the \texttt{Kernel} parameters were optimized with the grid search method while the \texttt{nu} with the method described in the text (step two of optimization).}
\end{table}

The second step is to find the optimal value for the parameter that sets the fraction of the population representing the instances the diagnostic will identify as normal instances (MS star-forming galaxies). 

In \texttt{oneclassSVM}, the parameter \texttt{nu} plays a pivotal role in determining the model's behavior. Acting as both an upper bound on the fraction of outliers and a lower bound on the fraction of support vectors, \texttt{nu} allows fine-tuning of the model's sensitivity to anomalies. By adjusting \texttt{nu}, we can achieve a balance between false positives and false negatives, influencing the model's classification of data points as normal or outliers.

This parameter, \texttt{nu}, needs to be optimized independently from the others because it is not related to the fitting process but rather to the trade-off between recall and contamination. Specifically, one could choose a boundary that includes all objects in the population (i.e., \texttt{nu} = 0, recall = 1). However, due to the continuous nature of galaxies, the transition between different activity classes occurs gradually. This means that selecting \texttt{nu} = 0 would result in high contamination, with other activity classes, such as AGN, being misclassified as star-forming galaxies.

After determining the optimal values of the parameters (all except \texttt{nu}) in the first step of the optimization process, we proceeded to find the optimal value of \texttt{nu} that simultaneously maximizes recall and minimizes contamination. To accomplish this, we utilized the sample of observed galaxies (see Sect. \ref{obs_spec}), aiming to tailor our diagnostic to perform optimally in real-world scenarios. We selected five values for \texttt{nu}, as presented in Table \ref{oneclassSVM_params}. The algorithm was trained iteratively, with \texttt{nu} being the only varying parameter while the others remained fixed at their optimal values from the first step of the optimization.

For each \texttt{nu} value, we calculated both recall, representing the fraction of correctly identified star-forming galaxies to the total star-forming galaxies, and contamination, representing the fraction of true AGN galaxies predicted as star-forming galaxies. Naturally, we anticipated both recall and contamination to increase as \texttt{nu} increased. As depicted in Fig. \ref{fig:opitmization}, both recall and contamination indeed increased with higher \texttt{nu} values. In the same figure, we observe a sharp increase in contamination above \texttt{nu} = 0.1. Based on these findings, we selected \texttt{nu} = 0.1 as it offers a balance between high recall and acceptable contamination.

\section{Mid-infrared spectral variations} \label{Appnxb}

The simulated spectra presented in Sect. \ref{simgalspec} are derived from five parameters (excluding SFH): the ionization parameter $(\log_{10}U)$, the PAH mass fraction $(q_{\rm PAH})$, the fraction of dust exposed to a power-law radiation field $(\gamma)$, the minimum starlight intensity heating the dust $(U_{\rm min})$, and the stellar metallicity $(\log_{10}Z/Z_\odot)$. To illustrate the impact of each parameter on the mid-infrared spectrum we perform the following exercise. To begin, we divide the parameters into two groups: the first includes parameters that have high impact on the mid-IR spectrum (i.e., $\gamma$, $U_{\rm min}$, and $q_{\rm PAH}$) while the second includes ones with low impact ($\log_{10}Z/Z_\odot$ and $\log_{10}U$).

For the first group of parameters, in order to ensure that the whole range of each parameter is explored, each time we select ten values from an equally spaced interval for the parameter we want to examine based on its 95\% confidence interval (see Table \ref{tab:physpar}). In addition, to ensure that the covariance between them is preserved, for each point selected above we draw random values for the other two parameters based on their probability distributions (Table \ref{tab:physpar}). Then, based on the above parameter grids we calculate the corresponding simulated spectra, by changing each time the parameter we study. These variations are presented in Fig. \ref{fig:var_params_dust}, demonstrating how each dust parameter, PAH dust fraction (top), dust fraction exposure (middle), and radiation intensity (bottom) affects the mid-infrared spectrum.

For each of the two low impact parameters we select ten equally spaced values based on their 95\% confidence interval (Table \ref{tab:physpar}) and we generate each spectrum by allowing on of the two to vary and keep all the rest parameters constant. Figure \ref{fig:var_params_gal} shows the variations caused by gas ionization parameter (top) and stellar metallicity (bottom). Upon comparing the two figures, it becomes evident that the parameters related to dust properties exhibit the most significant variation in the spectral region of interest (3-30\,$\mu$m).

\begin{figure}
\begin{center}
\includegraphics[scale=0.7]{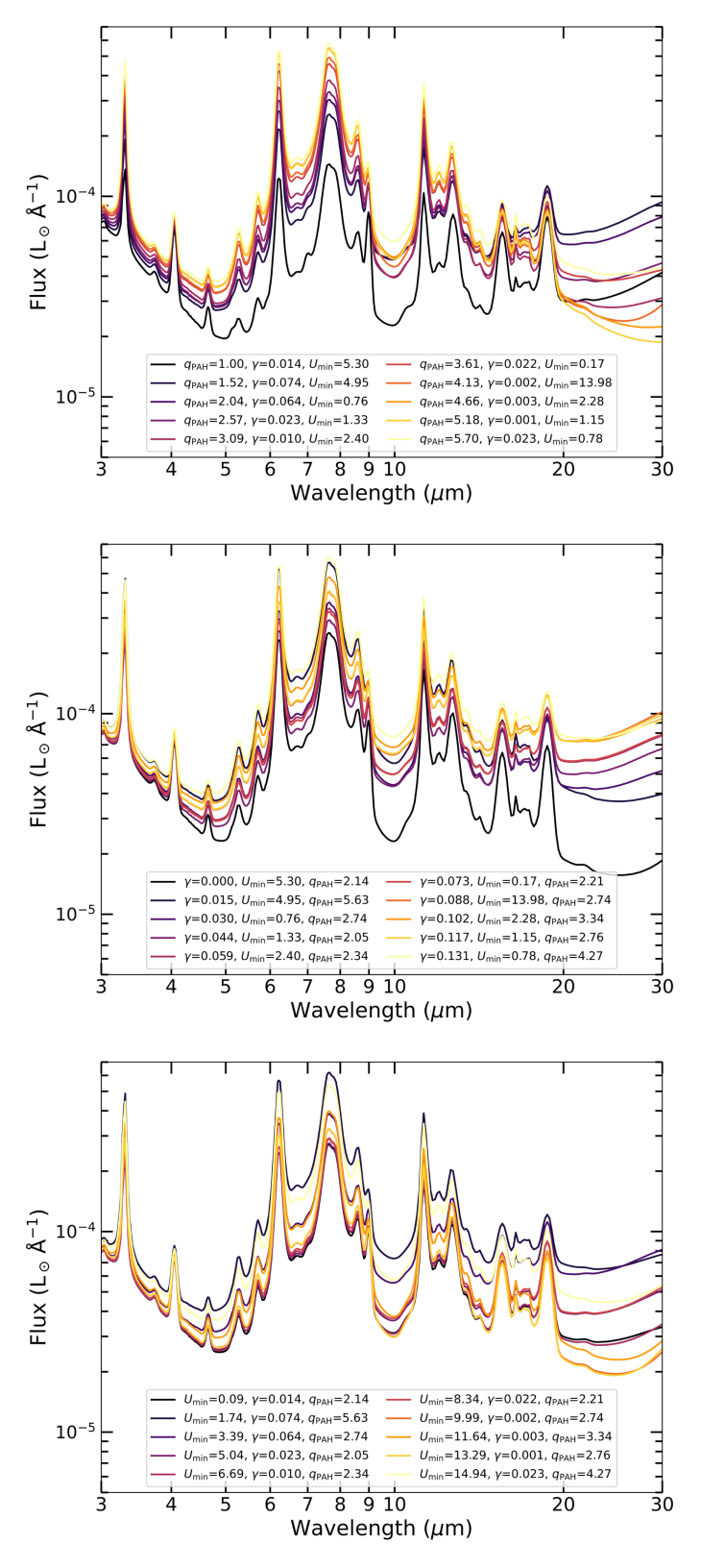}
\end{center}
    \caption{Synthetic mid-infrared (3–30\,$\mu$m) spectra demonstrating the influence of key dust parameters in the \cite{2007ApJ...663..866D} dust model. Each panel shows the variation of the spectrum by varying one parameter inside its 95\% confidence interval while the rest are drawn independent from their assumed distributions. Top: variation with the PAH mass fraction, $q_{\rm PAH}$. Middle: variation with the fraction of dust mass exposed to high radiation fields, $\gamma$. Bottom: variation with the minimum starlight intensity, $U_{\rm min}$. We see that these three parameters has significant impact on the shape of the PAH spectrum when compare to Fig. \ref{fig:var_params_gal}. All spectra have the same bolometric luminosity.}
\label{fig:var_params_dust}
\end{figure}

\begin{figure}
\begin{center}
\includegraphics[scale=0.45]{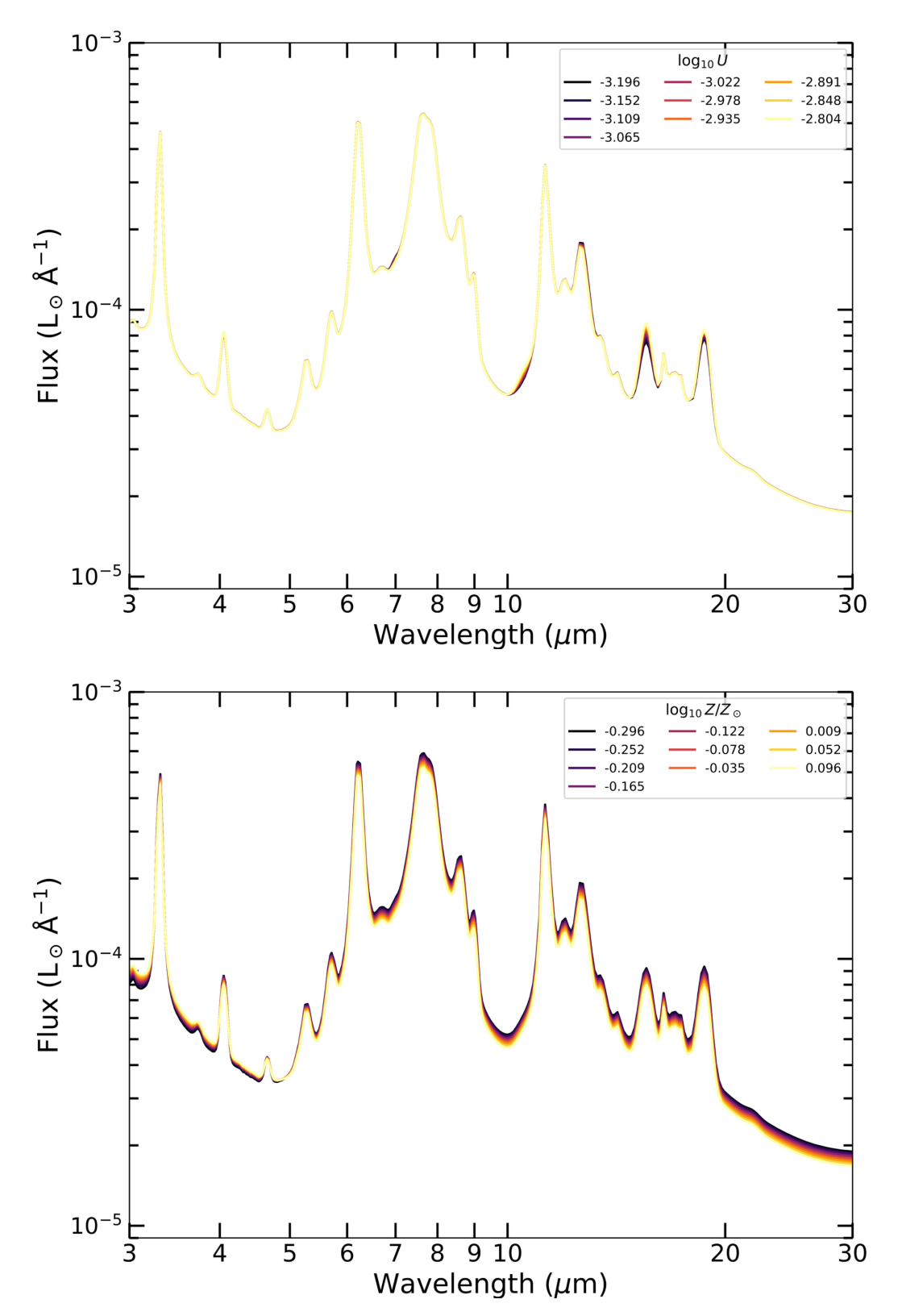}
\end{center}
    \caption{Synthetic mid-infrared galaxy spectra (3-30\,$\mu$m) highlighting the impact of ISM parameters. Top: variation of the spectrum as a function of ionization parameter $(\log_{10}U)$. Bottom: variation with gas-phase metallicity ($\log_{10}Z/Z_\odot$). Each line represents a mock galaxy spectrum generated with fixed ISM conditions while varying the labeled parameter. We see hat the impact of these parameters on the shape of PAH spectrum is subtle. All spectra have the same bolometric luminosity.}
\label{fig:var_params_gal}
\end{figure}

\end{appendix}
\end{document}